\documentclass[acmsmall,screen,nonacm]{acmart}

\usepackage[english]{babel}
\usepackage{lipsum}    
\usepackage{wrapfig}

\usepackage[normalem]{ulem}
\usepackage{fancyref}

\usepackage{tikz}
\usetikzlibrary{shapes,decorations.text,patterns}
\usepackage{adjustbox}

\usepackage{caption}
\usepackage{xspace}
\usepackage{subcaption}

\usepackage{tabularx}
\usepackage{algorithm}
\usepackage{algpseudocode}

\usepackage{verbatim}

\usepackage{graphicx}
\usepackage{multirow}

\usepackage[capitalise]{cleveref}

\usepackage{hyperref}

\crefname{section}{\S}{\S\S}

\renewcommand\footnotetextcopyrightpermission[1]{} 
\setcopyright{none}
\acmDOI{}
\acmISBN{}
\acmYear{}
\copyrightyear{}
\authorsaddresses{}

\newcommand{\showCODEN}[1]{\unskip}
\newcommand{\showDOI}[1]{\unskip}
\newcommand{\showLCCN}[1]{\unskip}
\newcommand{\showURL}[1]{\unskip}

\newcommand{\sectionref}[1]{$\S$\ref{#1}}

\def\spacehack{0}
\ifnum\spacehack=1
\newcommand{\fillgap}[1]{{\vspace{#1}}}
\else
\newcommand{\fillgap}[1]{}
\fi

\expandafter\def\expandafter\normalsize\expandafter{%
    \normalsize%
    \setlength\abovedisplayskip{1pt}%
    \setlength\belowdisplayskip{2pt}%
    \setlength\abovedisplayshortskip{-2pt}%
    \setlength\belowdisplayshortskip{2pt}%
}

\newcommand{\stt}[1]{{\small\texttt{#1}}}

\providecommand{\vs}{vs. }
\providecommand{\ie}{\emph{i.e.,} }
\providecommand{\eg}{\emph{e.g.,} }
\providecommand{\etc}{\emph{etc.} }
\providecommand{\sysname}{\textsc{Corn\-ifer}\xspace}

\providecommand{\myparab}[1]{\vspace{3pt}\noindent\textbf{#1} }
\newenvironment{packeditemize}{\begin{list}{$\bullet$}{\setlength{\itemsep}{0.2pt}\addtolength{\labelwidth}{0pt}\setlength{\leftmargin}{\labelwidth}\setlength{\listparindent}{\parindent}\setlength{\parsep}{0pt}\setlength{\topsep}{0pt}}}{\end{list}}
\title{\Large Cost-effective and performant virtual WANs with \sysname}
\author{Anjali}\affiliation{University of Wisconsin-Madison}\email{anjali@wisc.edu}
\author{Rachee Singh}\affiliation{Cornell University}\email{rachee@cs.cornell.edu}
\author{Michael M. Swift}\affiliation{University of Wisconsin-Madison}\email{swift@cs.wisc.edu}

\begin{abstract}
    Virtual wide-area networks (WANs) are WAN-as-a-service cloud offerings 
    that aim to bring the performance benefits of dedicated wide-area interconnects
    to enterprise customers. In this work, we show that the \emph{topology}
    of a virtual WAN can render it both performance and cost inefficient.
    We develop \sysname \footnote{https://github.com/multifacet/cornifer}, a tool that \emph{designs} virtual WAN topologies
    by deciding the number of virtual WAN nodes and their location in the
    cloud to minimize connection latency at
    low cost to enterprises. By leveraging millions of latency
    measurements from vantage points across the world to cloud points of presence,
    \sysname designs virtual WAN topologies that improve weighted client latency
    by 26\% and lower cost by 28\% compared to the state-of-the-art. 
    \sysname identifies virtual WAN topologies at the Pareto frontier
    of the deployment cost \vs connection latency trade-off and
    proposes a heuristic for automatic selection of Pareto-optimal
    virtual WAN topologies for enterprises.

\end{abstract}

\begin{document}

\maketitle

\section{Introduction}


\begin{wrapfigure}{r}{0.4\textwidth}
  \begin{center}
      \includegraphics[width=0.4\textwidth]{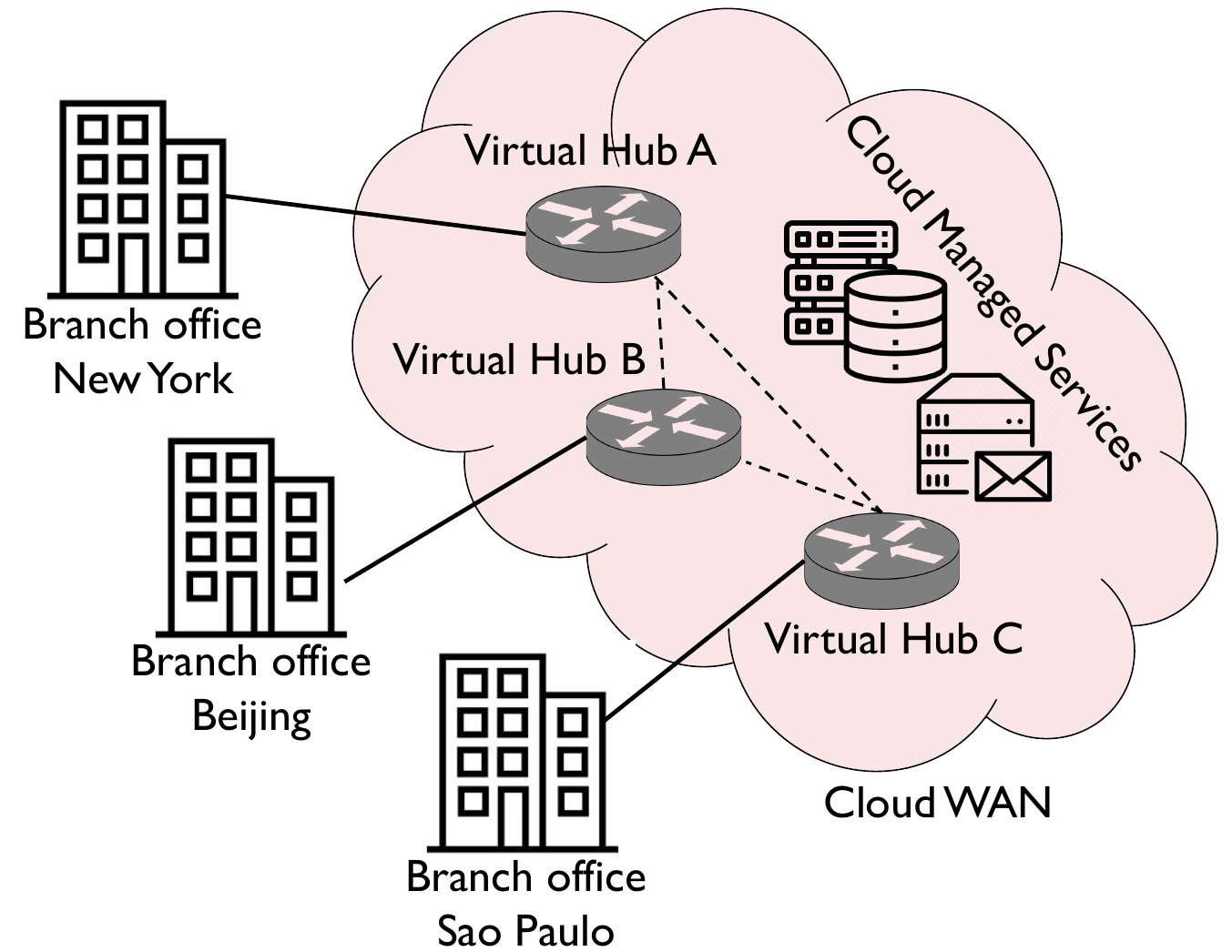}
  \end{center}
  \caption{Virtual WANs are cloud overlays. An example 
  topology consists of hubs $A$, $B$ and $C$ that are network gateways
  providing regional entry points into the virtual WAN.}
  \label{fig:vwan}
\end{wrapfigure}

Cloud providers have 
begun to offer \emph{virtualized wide-area networks} to 
enterprise customers~\cite{ref:magic-wan,ref:azure-vwan}.
Similar to virtualized compute in the cloud~\cite{ref:ec2}, 
virtualized WANs are \emph{WAN-as-a-service} offerings that 
bring the performance improvements of a dedicated WAN to enterprises
without significant infrastructural investment.
Fundamentally, enterprise virtualized WANs are \emph{network overlays}~\cite{10.1145/502034.502048} 
on the cloud that can provide fast connectivity between enterprise 
branch offices and their cloud-hosted resources (\eg mail, databases,
proprietary enterprise software). Figure~\ref{fig:vwan} shows an example 
enterprise virtualized WAN that connects three branch offices of 
a multi-continental enterprise and its employees. The overlay nodes, called \emph{virtualized WAN hubs} (\eg hubs $A$, $B$
and $C$ in Figure~\ref{fig:vwan}), are network gateways in the cloud,
providing regional points of connectivity into the cloud WAN.

Virtualized WANs promise improved network latency to cloud-hosted services
and between branch offices by using the cloud WAN instead of the 
public Internet for transport. However, 
these benefits depend on the virtual WAN's topology
\ie number and locations of WAN hubs. 
Today, enterprises build their virtual WANs by spawning 
virtual hubs in cloud datacenters (DCs) nearest to their branch
offices. As per this policy, the enterprise in Figure~\ref{fig:vwan} would 
place virtual hub $A$ in the New York DC, $B$ in the Beijing 
DC and so on. The intuition is that 
WAN hubs geographically closest to branch offices are most 
likely to offer the lowest latency entry 
into the cloud WAN. However, these intuitive designs 
can perform poorly in practice since both complex inter-domain and 
cloud intra-domain phenomenon impact the latency of an
enterprise virtual WAN.

\begin{wrapfigure}{r}{0.4\textwidth}
    \centering
    \begin{subfigure}{0.4\columnwidth}
      \includegraphics[width=\columnwidth]{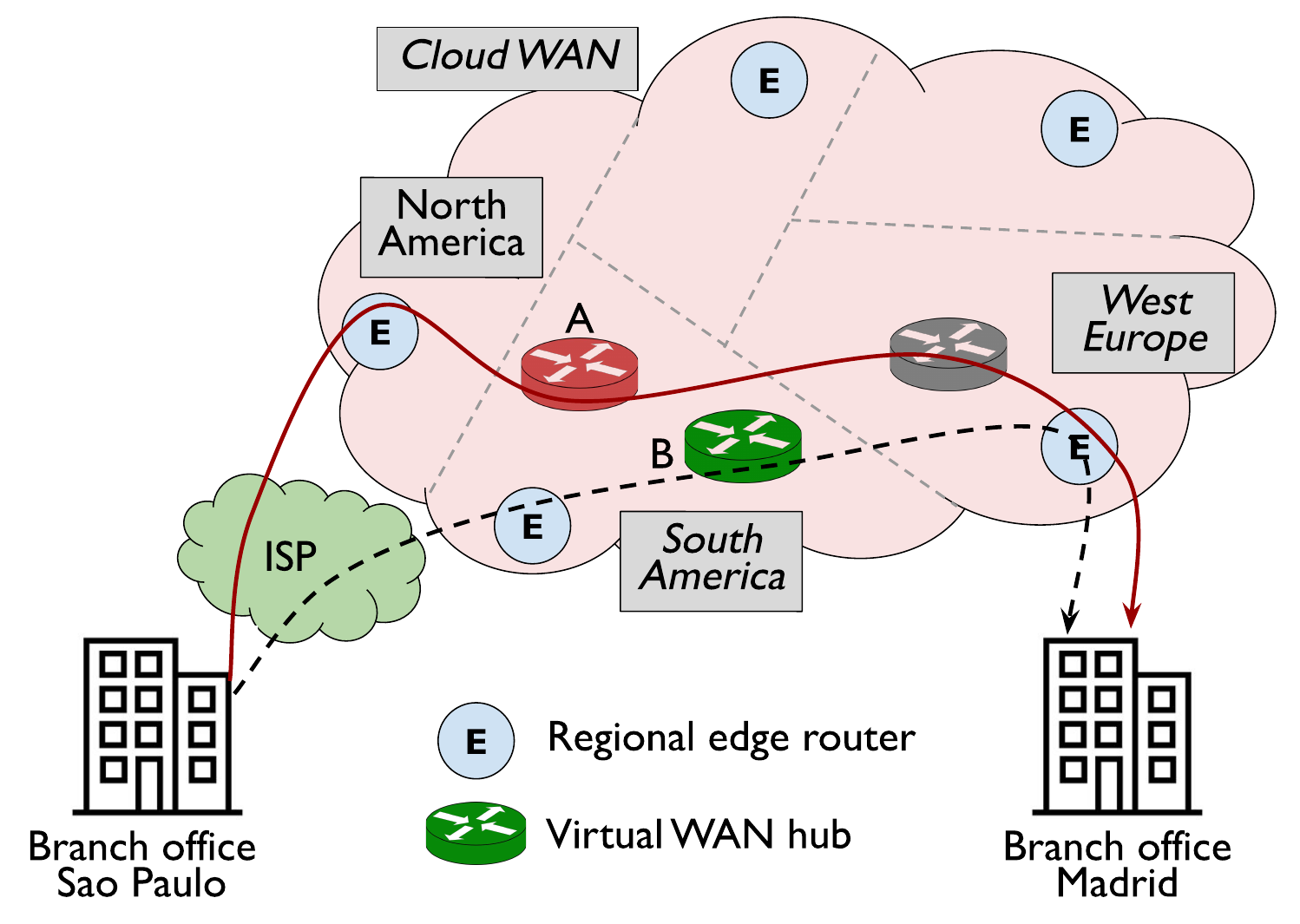}
       \caption{}
      {\vspace{0.1in}}
      \label{fig:vwan-bad}
    \end{subfigure}%
    \vfill
    \begin{subfigure}{0.4\columnwidth}
      \includegraphics[width=\columnwidth]{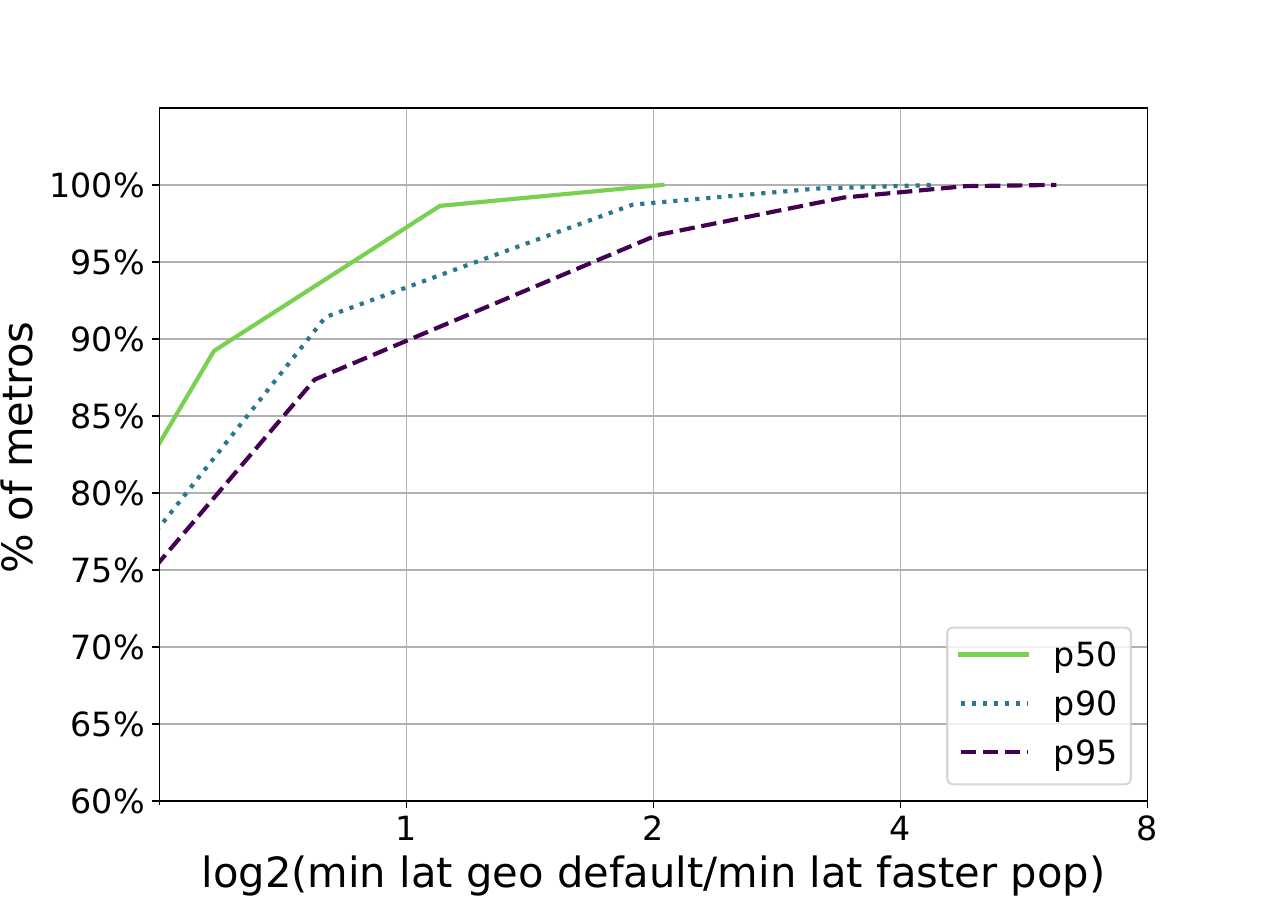}
      \caption{}
      \label{fig:latency_ratio}
       
     \end{subfigure} 
      \vfill
  
    {\vspace{0.15in}}
    \caption{~\ref{fig:vwan-bad} Problematic virtualized WAN topology. Dotted lines in the cloud
    network demarcate geographical regions (\eg S. America,
    N. America, W. Europe). ~\ref{fig:latency_ratio} shows 20\% of the metros take sub-optimal routes.}
  \end{wrapfigure}

\myparab{Sub-optimal virtualized WAN topologies are a reality.}
We observed significantly degraded performance 
from the virtual WAN of a large multi-continental 
enterprise (Figure~\ref{fig:vwan-bad}) due to inefficient 
hub placement in a commercial cloud network. The enterprise client 
spawned a hub ($A$) in the S. American region of the cloud WAN, 
nearest to their branch office in Sao Paulo.
However, the upstream Internet Service Provider (ISP) of the
Sao Paulo branch office routed the enterprise traffic to
an edge site in N. America instead of the edge site in 
S. America (\sectionref{sec:background}). This \emph{inter-domain} 
routing decision inflated client latency since their 
traffic was routed from S. America to N. America,
only to be routed back to the WAN hub in S. America before going to the WAN hub in Madrid. 
Furthermore, our analysis of a dataset containing latency measurements from metropolitan regions (metros) 
across the globe to different entry points, called points of presence (PoP), into a large commercial cloud WAN 
shows that about 20\% metros connect to sub-optimal routes (Figure~\ref{fig:latency_ratio}).

\myparab{Who to blame, ISP or the cloud provider?}
Ideally, the enterprise customer can resolve
the inflated latency problem in Figure~\ref{fig:vwan-bad}
by filing an incident with the cloud provider who can ask the upstream ISP to modify their routing policy
for $A$'s IP prefix. In fact,
the enterprise customer in Figure~\ref{fig:vwan-bad} pursued this 
action but it took several weeks for the problem to get resolved.
A cloud
provider could modify its own routing announcements towards 
hub $A$'s IP prefix to affect a policy change from ISPs. 
However, modifying prefix announcements is risky
since several cloud services are hosted in the same prefix.

\myparab{Hubs in alternate IP prefixes.}
For the example in Fig~\ref{fig:vwan-bad}, an alternate placement of the WAN hub for the Sao Paulo branch office
in the Lima datacenter ($B$) would have resolved this latency inflation 
since virtual WAN hubs in different datacenters are hosted in separate IP prefixes.
Therefore, while the enterprise's upstream 
ISP routes sub-optimally to hub $A$'s IP prefix via North America,
hub $B$'s IP prefix is not affected by that routing decision.

\myparab{Designing virtual WANs around routing decisions.}
We develop \sysname, a tool that enables
enterprise clients to \emph{independently} deploy
virtual WAN topologies that achieve the best performance possible,
taking the inter- and intra-domain routing behavior as a given. 
Designing performant virtual WAN topologies is challenging in practice.
First, designing topologies that avoid inflated inter-domain latency
requires a \emph{global view} into the \emph{dynamic} network performance from 
client locations to the cloud WAN. Second, the cost of virtual 
WAN deployments scales linearly with the size of the topology~\cite{ref:vwanpricing}.
Naive designs can reduce the client latency by a
placement of hubs in all datacenters of the cloud WAN. However, such 
deployments are expensive. Finally, reliable data sources capturing the  
global view of dynamic latency measurements 
are needed to design the correct topologies, but are not widely
available. 


\myparab{Contributions.} We make the following contributions in this work: 

\begin{packeditemize}
  \item \myparab{Identify opportunities to improve virtual 
  WAN performance.} We analyze a dataset of latency measurements 
  from metros across the globe to different PoPs, and show that 
  sub-optimal routing causes inflated latency for roughly 20\% of 
  the metros in our dataset (\S\ref{sec:quantify}).
  \item \myparab{Novel optimization formulation.} We formulate 
  the problem of placing hubs in a virtual WAN as an optimization. 
  We build \sysname, a tool that optimizes latency (\emph{performance-optimal}), 
  cost (\emph{cost-optimal}), or both (\emph{Pareto frontier}) 
  when making hub placement decisions (\S\ref{sec:magpie}). 
  In the \emph{cost-optimal} mode, \sysname designs virtual WAN 
  topologies that use 28\% fewer WAN hubs than state-of-the-art techniques. 
  In \emph{performance-optimal} mode, \sysname improves weighted latency 
  of enterprise clients by 26\%. \sysname balances the goals
  of performance and cost by computing the Pareto frontier of 
  optimal solutions. Virtual WAN designs at the 
  Pareto frontier improve weighted client latency by 22\% at 25\% 
  lower cost than state-of-the-art. \sysname can design optimal 
  virtual WANs for enterprises with hundreds of branch offices within 
  a few seconds (\S\ref{sec:evaluation}).
  \item \myparab{Adapting anycast routing data as input to the optimization.} We demonstrate how to use 
  the latency performance data of anycast routing towards cloud providers, which are partial, noisy, and 
  spotty, for robustly making hub placement decisions (\S\ref{sec:quantify}).
  \item \myparab{Maintenance and reconfiguration.} We show that \sysname's placements are stable and provide benefits 
  for extended periods. However, recomputing placements every few days with a fast-moving average
  can keep latency low as internet conditions change (\S\ref{sec:maintain}).
\end{packeditemize}

\myparab{Ethics.} This work does not raise any ethical concerns.

\section{WAN-as-a-service}
\label{sec:background}

Many enterprises (\eg banks, retail companies) have geo-distributed
footprints with branch offices and stores across the world. These
offices often connect to the same global database to keep track of
customer accounts, sales, \etc The bulk of the local infrastructure at
branch offices has migrated to the cloud and enterprises have
smaller footprints \emph{on premises} to access cloud-hosted
applications like databases, mail servers, \etc Moreover,
branch offices communicate and exchange data with
each other. The two types of enterprise communication, branch to cloud
services and inter-branch used the public Internet, which is known to
 be less reliable and worse in performance than
dedicated WANs operated by large cloud and Internet service
providers. However, private WANs are billion-dollar assets. To bring the
benefits of a dedicated WAN to more enterprises, cloud companies offer
\emph{virtual WANs}~\cite{ref:magic-wan, ref:azure-vwan,
  ref:cisco-iwan, ref:aws-cloud-wan}. Virtual WANs are
\emph{WAN-as-a-service} offerings that allow enterprises to leverage
the cloud WAN as the interconnect between their branch offices
(Figure~\ref{fig:vwan}).

\subsection{Cloud Virtual WANs}
Virtual WANs are network overlays on the cloud~\cite{ref:ron}. Clients connect to
nodes of the overlay  called \emph{virtual WAN hubs}, which
are network gateways that offer regional points of connectivity
into the cloud. When spawning a virtual WAN deployment, enterprises
decide the number of hubs and their locations in the cloud WAN.
Virtual networks directly connected to the virtual WAN hubs host 
cloud-managed service like database and application servers.

\myparab{Virtual hubs.}
Virtual WAN hubs are software gateways that run in DCs.
Traffic from enterprise branch offices to cloud-hosted 
services or inter-branch office traffic transits one 
or more virtual WAN hubs. 
Enterprises create hubs by specifying the cloud region 
where the hub will be hosted. In reality, hubs reside in DCs in the specified region. We assume that cloud providers allow enterprises to specify the
datacenter where a hub is placed.

\myparab{Hub connection limits.}
There are limits on how many client connections a virtual WAN hub 
can sustain. 
The cost of the virtual WAN to the 
enterprise customer is a linear function of the number of WAN hubs, their
uptime and the volume of data transferred between the hubs~\cite{ref:vwanpricing}.

\myparab{Virtual WAN topology.}
By design, virtual WAN hubs in the same virtual WAN deployment form 
a fully connected graph. Therefore, the number of virtual WAN hubs and their
location in the cloud datacenters completely defines a virtual WAN's
topology. For example, Figure~\ref{fig:vwan} shows a virtual WAN topology with three
hubs $A, B, C$.

\subsection{Virtual WAN performance}
Virtual WANs aim to enable lower
latency access to cloud-managed services and between enterprise branch offices 
compared to the public Internet. 

\myparab{Anycast routing to cloud PoPs.}
Virtual WAN hubs are hosted in IP address
space owned by the cloud provider. 
The IP space may also host other cloud services, and its prefixes are 
announced to the Internet by BGP at multiple cloud 
PoPs and are said to be \emph{anycasted}~\cite{ref:matt-imc}. 
ISPs can select any path to reach an anycast prefix;
BGP's best-path selection process guides path
selection. Anycast routing has been shown to perform well 
by drawing traffic to PoPs nearest to the client~\cite{ref:matt-imc}.

\begin{wrapfigure}{R}{0.4\textwidth}
  \centering
  \includegraphics[width=0.4\textwidth]{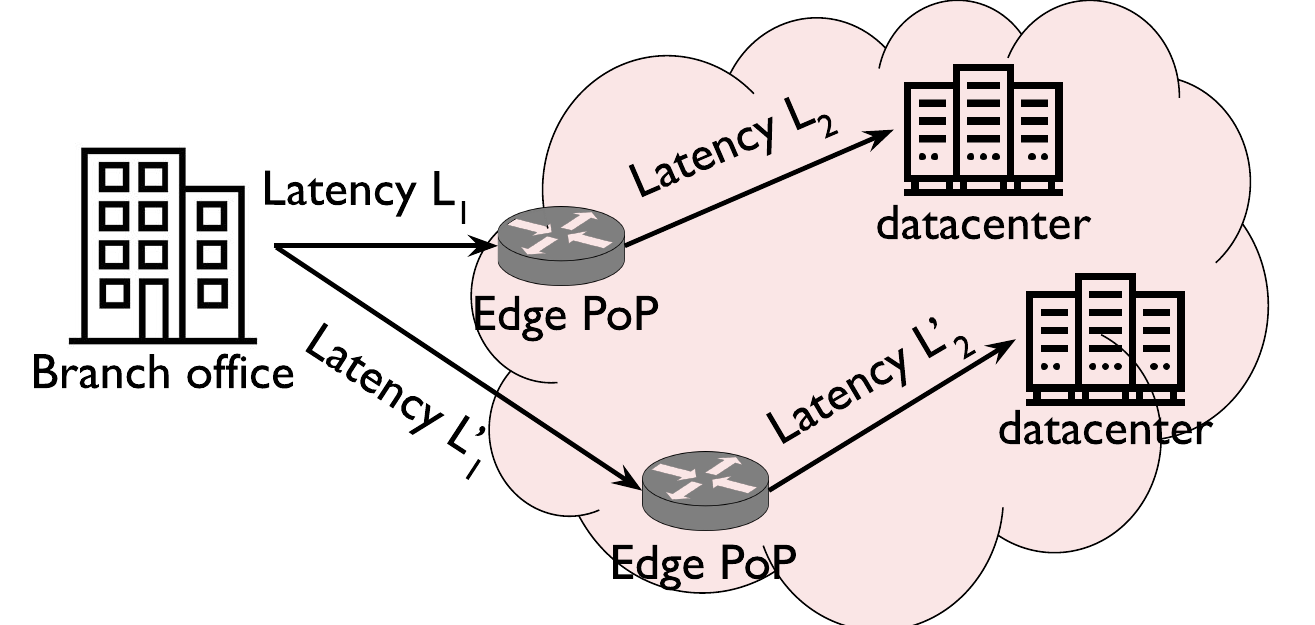}
  \caption{Inter-domain and intra-domain latency from enterprise
  branch offices to virtual WANs. $L_2$ and $L_2\prime$ measure the
  latency from edge PoP to the nearest datacenter.}
  \label{fig:lat_intra_domain}
\end{wrapfigure}

\myparab{Latency components.} There are two components of
virtual WAN connection latency (Figure~\ref{fig:lat_intra_domain}):
Traffic from branch offices traverses the public Internet until it
reaches a cloud PoP (shown as $L_1$ and $L_1\prime$ in
Figure~\ref{fig:lat_intra_domain}). This portion of the path
contributes the inter-domain latency of virtual WAN connections 
and depends on the path selected by ISPs. Once the virtual WAN 
traffic reaches the cloud edge, it is routed to
the DC hosting the virtual WAN hub ($L_2$ and
$L_2\prime$ in Figure~\ref{fig:lat_intra_domain}). This latency
depends on intra-domain routing decisions made by the cloud network.

\myparab{Inter-domain latency is over 80\% of the connection latency.}
We measure the latency from hundreds of client locations to the 
edge PoPs of a large commercial cloud provider (\S\ref{sec:quantify}).
We compute the median latency over hundreds of latency measurements in a day
from client locations to cloud edge PoPs and on the same day measure ping latencies from cloud edges 
to all datacenter gateway routers in the cloud. 
These measurements show that the latency 
to reach the cloud edge ($L_1$ or $L_1\prime$ in Figure~\ref{fig:lat_intra_domain})
is 80\% of the overall latency to the edge and subsequently the nearest datacenter in 
over 75\% of the cases. In other words, $\frac{L_1}{L_1 + L_2}$ is
$\ge 80\%$. Thus, we focus on the inter-domain
portion of client latency to improve the performance of virtual 
WANs. 
Traffic between branch offices additionally traverses 
inter-datacenter links in the cloud, and  SDN-based performance-optimal 
routing implemented in cloud providers for intra-domain 
traffic~\cite{ref:swan, ref:b4} keeps this latency minimal. 

\myparab{Performance anomalies.}
In some cases, ISPs route traffic towards anycast prefixes through
longer paths on the Internet because of  
traffic engineering policies in the ISP network, attempts to load balance
across multiple network links or even configuration errors. As a result,
branch clients observe inflated latencies while accessing cloud
services.

\myparab{Alternate placement of hubs can fix performance.}
If the enterprise in
Figure~\ref{fig:vwan-bad} had placed their hub in a different DC 
in the same region \ie S. America, their traffic would be \emph{routed around}
the performance anomaly in Figure~\ref{fig:vwan-bad}. This is because 
virtual WAN hub deployments in different parts of the cloud WAN use 
different IP address spaces. Thus, the sub-optimal route 
to hub $A$'s prefix by the upstream ISP of the branch office does not 
affect the route to hub $B$'s prefix. The branch office, while being 
geographically closest to hub $A$, may receive much better performance if 
it routes traffic through hub $B$.

\subsection{Designing virtual WANs}
In this work, we leverage the insight that alternate placement of virtual WAN
hubs can alleviate the effect of inter-domain performance anomalies in virtual WAN
connection latency. Before formulating the problem of designing performant 
virtual WAN topologies, we describe how enterprises spawn their virtual WAN
topologies today.

\myparab{Popular heuristics.} The main guideline enterprises use to
design virtual WANs is physical distance. Assuming proximity as a
crude proxy for network latency, enterprises often decide the location
of virtual WAN hubs based on proximity to the geographic locations of
their branch offices. Apart from lower end-to-end connection latency, enterprises 
aim for low cost virtual WAN deployments. The cost of a virtual WAN deployment is a 
linear function of the number of virtual WAN hubs in the topology, so enterprises
aim to reduce the number of hubs in their deployments.

\myparab{Tradeoffs.}
The brute-force solution of the virtual WAN design problem 
places one or more hubs for every branch office in DCs at the
shortest latency from the branch office. This design can be 
prohibitively expensive due to a large number of hubs in the deployment.
Ideally, the solution to the virtual WAN design problem should
explore the \emph{Pareto frontier} of cost \vs performance tradeoff, allowing
enterprises to get the best performing topology at the lowest cost.

\section{\sysname Overview}
\label{sec:magpie}

\begin{wrapfigure}{R}{0.4\textwidth}
  \centering
  \includegraphics[width=0.4\columnwidth]{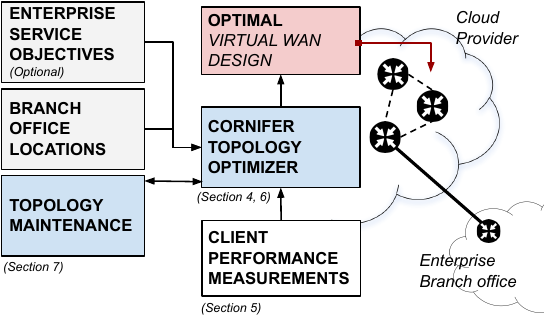}
  \caption{\sysname Design.}

  \label{fig:magpie-design}
\end{wrapfigure}

We develop \sysname, a tool that designs performance and cost-optimal 
virtual WAN deployments for enterprises. An enterprise provides
the geographical footprint of their branch offices and employees 
to \sysname as input. \sysname leverages a streaming
latency measurement pipeline (\S\ref{sec:quantify}) to profile the
performance of connections to cloud PoPs from client metros 
\ie the inter-domain portion of virtual WAN connections 
described in \S\ref{sec:background}. 
It recommends the number and datacenter locations of virtual WAN hubs 
for the enterprise. 
Figure~\ref{fig:magpie-design} shows the design of \sysname.

\subsection{Design goals}
\sysname has two design goals: 
(1) Placing WAN hubs into topologies that \emph{minimize the latency} across 
client connections. (2) Use the \emph{fewest hubs} to
minimize latency, in turn reducing the cost of the virtual WAN deployment. 

Enterprises can configure \sysname in
\emph{performance-optimal} or \emph{cost-optimal} modes. The performance-optimal
mode minimizes the weighted latency of the WAN topology with no 
upper bound on the cost. The cost-optimal mode reduces 
the weighted latency using a virtual WAN deployment of near-minimal cost.
Both modes produce hub placements based on the optimization
goal(s). Alternatively, an enterprise can trade off the two objectives
by exploring the cost/performance
Pareto frontier.

\subsection{Design decisions}
We describe the decisions that led to the design of \sysname.

\myparab{\sysname's interface to enterprises.}
\sysname seeks two inputs from enterprise clients: (1) the location
of enterprise branch offices, and (2) the maximum number of client connections
from each branch office. 
For instance, an enterprise can specify  
three branch offices, 
each with 500 people. 

\myparab{Connection latency to edge PoPs instead of DCs.}
While WAN hubs are hosted in DCs, it is sufficient for \sysname
to minimize the connection latency to the 
cloud edge PoP instead of the DC since
inter-domain latency dominates the total connection latency to 
the WAN hub (\S\ref{sec:background}).

\myparab{\sysname's optimal hub placement.}
Equipped with the enterprise inputs and performance measurements from the 
cloud provider, \sysname formulates the optimization problem as a mixed integer 
linear program (MILP). The decision variables 
decide how many and which PoPs should be selected for WAN hubs for
one or more branch offices (\S\ref{sec:hub_placement}).

\myparab{Output.} \sysname builds optimal virtual WAN topologies
for the enterprise and outputs the number 
and datacenter locations of virtual WAN hubs in the suggested topology.
Since all hubs in a topology are fully connected, the number
and locations fully specify a virtual WAN topology.

\myparab{\sysname needs global measurements.}
\sysname optimization algorithm needs global latency measurements from
client locations to cloud edge PoPs. We were given access
to such a dataset by a large commercial cloud provider. Other cloud 
providers can use existing systems that measure client to cloud edge latency~\cite{ref:odin} 
to provide inputs to \sysname. Despite the availability
of this dataset, we had to contend with a number of challenges of using
this data for \sysname (\S\ref{sec:quantify}). 

\myparab{\sysname's deployment.}
\sysname can be deployed as a built-in tool by cloud provider with 
WAN-as-a-service offerings. In such cases, \sysname can suggest hub
deployments to enterprises in a similar way as cloud provider 
online portals~\cite{ref:azportal} suggest virtual machine sizes to customers.

\section{Optimal Virtual WAN topologies}
\label{sec:hub_placement}

\begin{wraptable}{r}{.5\textwidth}
    \begin{tabular}{ll}
    \multicolumn{2}{c}{\textbf{Inputs:}} \\
    $P$: & set of cloud PoPs, $|P| = p$ \\
    $\beta$: & max. connections per virtual WAN hub \\
    $L_{ij}$: & measured latency from metro $j$ to PoP $i$ \\
    $M$: & branch office metro locations, $|M| = m$ \\
    $C_j$: & max. connections from branch office $j$\\
    $K$: & max. number of unique PoPs \\[1em]
\end{tabular}

\begin{tabular}{ll}
    \multicolumn{2}{c}{\textbf{Outputs:}} \\
    $u_i$: & indicator; 1 a hub is placed in PoP $i$ \\
    $U_{ji}$: & indicator; 1 if PoP $i$ selected for metro $j$ \\
    $k_{min}$:& min. unique PoPs with feasible solution  \\
    $k_{max}$:& max. unique PoPs with feasible solution  \\[1em]
\end{tabular}

\begin{tabular}{lll}
    \multicolumn{3}{c}{\textbf{Minimize:}} $ \sum_{j=1}^{m} C_j \cdot \sum_{i=1}^{p} U_{ij} \cdot L_{ij} $\\
    (1)       &    $ u_{i} \in \{0, 1\}$                             &    $\forall i \in P$ \\        
    (2)     &    $ U_{ji} \in \{0, 1\}$                            &    $\forall j \in M, i \in P$ \\
    (\ref{eq:onepop-metro})   &    $\sum_{i=1}^{p}  U_{ji} = 1$                      &    $\forall j \in M$ \\
    (\ref{eq:unique-pop})     &    $u_{i} - \sum_{j=1}^{m} U_{ji} \geq 0$            &    $\forall i \in P$ \\
    (\ref{eq:num-hubs-bound}) &    $\sum_{i=1}^{p}  u_{i} <= K$                      &    \\[0.6em]
\end{tabular}

\caption{\sysname's algorithm for virtual WAN design (Algorithm 1)}
\label{alg:cap}
\end{wraptable}

Designing virtual WAN topologies entails deciding the number of
hubs and their locations in the cloud. While optimization has 
been applied to overlays before\cite{ref:skyplane}, 
we develop a novel optimization 
formulation tailored to the 
virtual WAN design problem space.
\sysname's optimization algorithm takes three types of inputs:
(1) inputs from the cloud platform that offers to host virtual WANs (2)
inputs from live measurements of connection latency from clients to the cloud
and (3) inputs from the enterprise designing their virtual WAN deployment.

\myparab{Inputs from the cloud provider.}
$P$ is the set of cloud PoPs adjacent to datacenters that host
virtual WAN hubs. There is a constant limit on the number of simultaneous
connections to a virtual WAN hub in cloud networks regardless of hub
location. We represent the per-hub connection limit with ${\beta}$ and set ${\beta}$ to 1000~\cite{ref:vwan_faq}
in our evaluation (\S\ref{sec:evaluation}).

\myparab{Inputs from live measurements.} \sysname needs 
latency measurements between branch offices to cloud edge PoPs.
We represent the latency from metro $j$ to cloud PoP $i$ as matrix $L_{ij}$.
Collecting this data at planet-scale is challenging. We leverage 
systems for collecting similar data developed in previous work~\cite{ref:odin}.
The data consists of periodic latency measurements from metros to PoPs. 
We aggregate the latency between a pair of metro and PoP
to a percentile (\eg median) of the latency distribution in a window. The enterprise
customer's desired latency dictates which percentile of the
latency distribution is input to \sysname. We note that a full
cross-product of latency between client's and PoPs is not
needed since many PoPs are too distant
from a client metro to be considered.
We grapple with the challenges of using this data meaningfully
with \sysname in \S\ref{sec:quantify}. 

\myparab{Inputs from the enterprise.}
The optimization algorithm takes the location of enterprise branch offices
at the granularity of a metropolitan location. $M$ is the set of all branch
office locations. The enterprise customer also provides the number of 
expected connections from each branch office, represented by 
the vector $C$ of size $m$. We assume that each branch office connects 
to one virtual WAN hub (\ie, it does not split traffic across multiple
hubs).

\myparab{Decision variables of the optimization.}
We need decision variables to count the number 
of unique PoPs selected by \sysname for placing hubs. For this, we 
introduce the indicator output variable $u_i$:
$u_i$ is $1$ if \sysname decides to place a WAN hub in PoP $i$.
To ensure that only one PoP is selected to place 
a hub for a client metro location, we use another indicator decision variable, $U$:
$U_{ji}$ is $1$ if the hub for metro $j$ is placed in PoP $i$. 

\myparab{Constraints.}
Both $u_i$ and $U_{ji}$ are indicator variables and can
only take values $1$ or $0$ (constraints (1) and (2) in Alg.~\ref{alg:cap}).

\noindent We ensure that connections from one client metro 
go to only one hub placed in a PoP:

 \begin{equation} \tag{3}
     \sum_{i=1}^{p}  U_{ji} = 1,  \forall j \in M
 \label{eq:onepop-metro}
 \end{equation}

 \noindent We ensure that only one PoP is selected to place the hub
 for one client metro location:
 \begin{equation}  \tag{4}
 u_{i} - \sum_{j=1}^{m} U_{ji} \geq 0
 \label{eq:unique-pop}
 \end{equation}
 \vspace{-1em}

\sysname designs virtual WAN topologies within ${K}$
PoPs. We search for the smallest ${K}$ for which the
optimization can find a feasible solution. From there, we
backtrack ${K}$ in steps of 1. This allows us to explore the 
cost \vs performance Pareto frontier. 
The total unique PoPs selected for placing WAN hubs
should not exceed the threshold ${K}$
(only enforced for \emph{cost-optimal} mode of \sysname):
 
 \begin{equation}  \tag{5}
     \sum_{i=1}^{p}  u_{i} <= K
 \label{eq:num-hubs-bound}
 \end{equation}
 \vspace{-1em}

We set this value so that the optimization achieves the objective with
a minimum number of PoPs. This reduces the number of
hubs as metros with the same PoPs can be merged, aggregating the
connections. To compute the number of hubs, we sum the
connections to each PoP across all metros that chose it and
divide those sums by $\beta$ to get the total number of hubs required
at each PoP. Adding these yields the total number of required
hubs for the topology.

\myparab{Objective Function.}
\sysname minimizes the sum of latencies from each branch office to
the PoP it is mapped to, weighted by the number of client connections from
that branch. The optimization finds PoPs ($u_i$)
that minimize the weighted latency while ensuring that all the constraints are met. 
Moreover, for \emph{cost-optimal} mode we re-run the algorithm for different values of ${K}$, to find
the least possible ${K}$ that gives a feasible solution. This outputs $k_{min}$.

\myparab{Operation modes.}\sysname runs in two modes, 
\stt{l\_optimal}, and \stt{k\_optimal}. \stt{l\_optimal} is 
the \emph{performance-optimal} mode, and ignores constraint~\ref{eq:num-hubs-bound},
optimizing only for weighted latency. This mode uses as many hubs as
necessary for the lowest latency, which we term as $k_{max}$. Adding hubs beyond 
$k_{max}$ either gives an infeasible solution or does not improve latency. 

\stt{k\_optimal} is the \emph{cost-optimal} mode, and minimizes the weighted 
latency with the fewest of PoPs and outputs $k_{min}$. 
$k_{max}$ and $k_{min}$ are the two extreme ends of the Pareto frontier for cost.

\myparab{Pareto Frontier.}
\sysname explores all possible solutions on cost vs. performance Pareto frontier
by solving the objective at each value of $K$ between $k_{min}$ and 
$k_{max}$. These solutions are virtual WAN topologies
with minimum weighted latency at all useful $K$ values.

We note that the least value of $K$ would be $1$ if \sysname had 
access to latency measurements between the full cross-product of PoPs 
and metros. However, \sysname is limited by the availability of 
global latency measurements due to the dynamics of anycast on the Internet.
Anycast-based latency measurements may not cover all the possible PoP
and metro pairs. This limits the possible minimum ${K}$ values that can give 
a feasible solution to the optimization. Finally, while \sysname's 
post-processing step is near-optimal, it may not be optimal in cases
where metros with fractional use of a hub will add an extra hub
nearby rather than share a hub further away. Algorithm~\ref{alg:cap} summarizes the \sysname's optimization formulation 
using equations (1)-(6).

\section{Challenges of real-world data}
\label{sec:quantify}

We describe the real-world dataset that we 
use to evaluate \sysname and quantify the potential of improving 
latency of virtual WAN deployments by showing the existence of better
alternate locations for virtual WAN hubs.

\myparab{Dataset.} We leverage an existing pipeline of latency measurements
performed globally from client endpoints towards different cloud WAN PoPs~\cite{ref:odin}.
We aggregate the measurements to the granularity of client metro
locations as sources and cloud PoPs as destinations.
We analyze hundreds of millions of measurements per month to 
PoPs, spanning hundreds of metros, 
each with thousands of latency measurements on any given day. 
We use data collected from April -- July, 2021.
To make the dataset uniform, we aggregate the samples 
to the median, 90$^{th}$ and 95$^{th}$ percentile latency in each ten 
minute time window between a metro and a PoP.
One row in the dataset contains: client metro location and network,
timestamp, PoP, median, 90$^{th}$, and 95$^{th}$ percentile latencies
(in milliseconds) for the ten minute window. 

\begin{wraptable}{r}{0.43\textwidth}
  \centering
  \small
    \begin{tabular}{|l|r|}
    \hline
    Total samples                                                                        & 9,362,717                      \\ \hline
    \begin{tabular}[c]{@{}l@{}}Total samples of metros \textgreater 1 PoP\end{tabular} & \multicolumn{1}{l|}{7,206,584} \\ \hline
    Metros with 1 PoP                                                                    & 166                            \\ \hline
    Metros \textgreater 1 PoP(s)                                                         & 154                            \\ \hline
    \end{tabular}
    \caption{Summary of dataset.}
    \label{tab:analysis_data}
\end{wraptable}

\myparab{Representativeness of the data.}
Our dataset is representative of real client latencies to PoPs for two
reasons. First, these measurements are gathered from 
the transfer of a small file from the PoP to the client location. 
Data-transfer based measurements better estimate the latency of client-to-cloud communication than mechanisms
that use ICMP packets because they are often treated differently from data packets
by routers. Second, the set of clients performing these measurements 
are already accessing services in the cloud network we study, making
the measurement client-metro set an accurate reflection of the actual 
clients of the cloud.

\myparab{Limitations of the dataset.}
The measurements from clients in our dataset are \emph{organic} \ie
a measurement occurs when clients of the commercial cloud provider
access a popular cloud-hosted website. The measurement infrastructure records
the client's metro while using client-side code to
measure to cloud edge PoPs~\cite{ref:odin}. As a result, 
the measurements do not occur uniformly across time or metros. 
The measurement methodology relies on anycast routes to PoPs, 
similar to real-world customer traffic. As a result, latency measurements 
from a client metro are directed towards a small set of cloud edge PoPs --- 
decided by anycast routing. This means that the dataset 
we use does not have a full cross-product of measurements from all 
client metros to all PoPs.

\myparab{Measurements to default PoPs.}  Our dataset captures latency
measurements to different PoPs from every client location. This
includes the PoP chosen by anycast routing as well as a few alternates.
Anycast routing steers traffic to what
we call the \emph{anycast-default} PoP. The measurements to alternate PoPs in
our dataset are enabled using special configuration at the cloud
PoP~\cite{ref:odin} router. However, the measurements do {\em not} record
which PoP is the \emph{anycast-default}. To identify the \emph{anycast-default}
PoP for a client metro, we use the count of measurements
towards all PoPs from a given metro. For most locations, one
PoP receives significantly more measurements, indicating it is used
more. We term the PoP with the most measurements for
a location as the \emph{anycast-default} PoP. 
Per prior work~\cite{ref:matt-imc}, for $\approx$90\% of 
client prefixes, anycast connects to the geographically closest PoP. 
We also use coordinates of PoPs in our dataset to compute the 
geographically closest PoP for all client metros, which we term the \emph{geo-default} PoP.

\begin{figure*}[t]
    \centering
            \begin{subfigure}[t]{0.32\textwidth}
             \centering
             \includegraphics[width=\columnwidth]{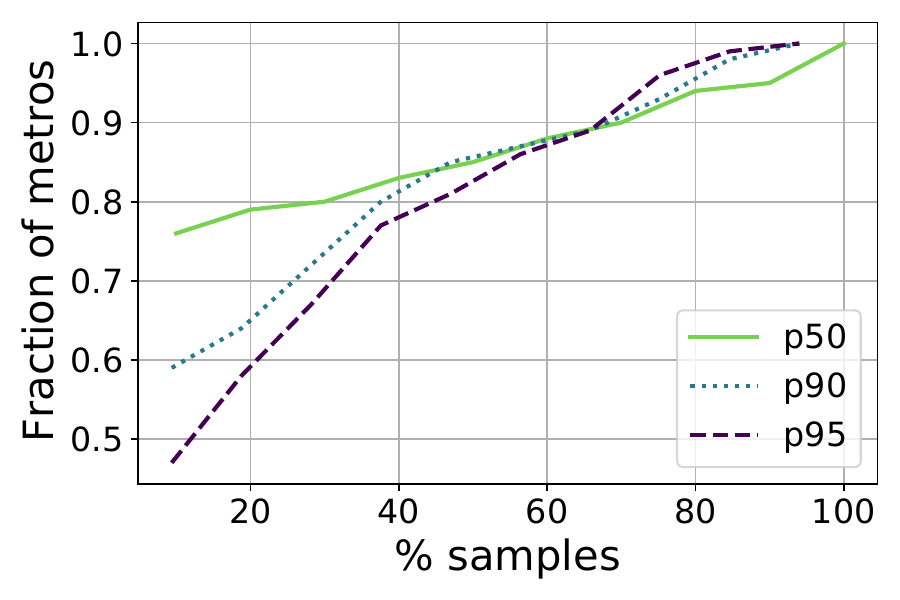}

             \caption{Faster PoP \% (10 minutes).}
             \label{fig:non_default_count}
         \end{subfigure}
         \hfill
         \begin{subfigure}[t]{0.32\textwidth}
             \centering
             \includegraphics[width=\columnwidth]{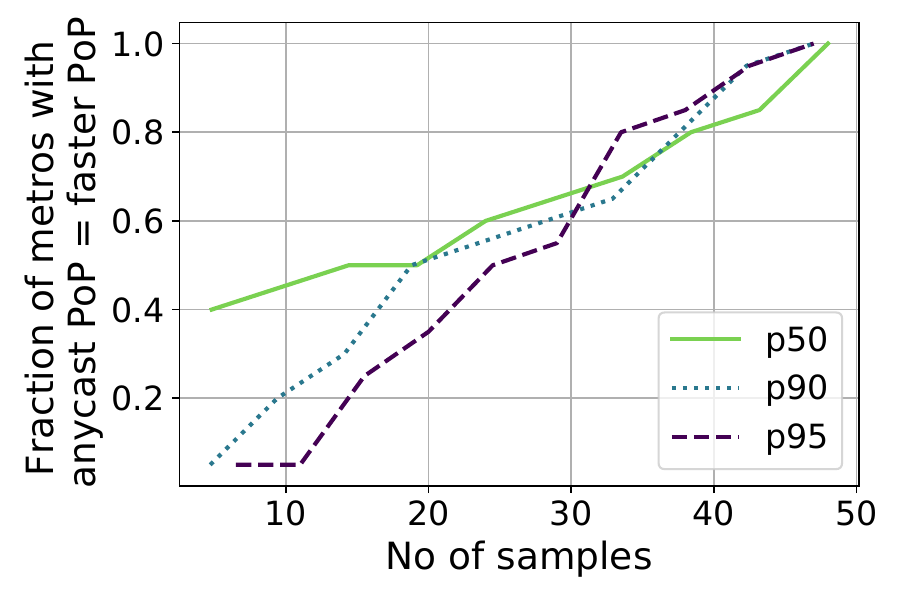}
             \caption{Faster PoP same as anycast.}
             \label{fig:anycast_non_default_count}
         \end{subfigure}
         \hfill
         \begin{subfigure}[t]{0.32\textwidth}
             \centering
             \includegraphics[width=\columnwidth]{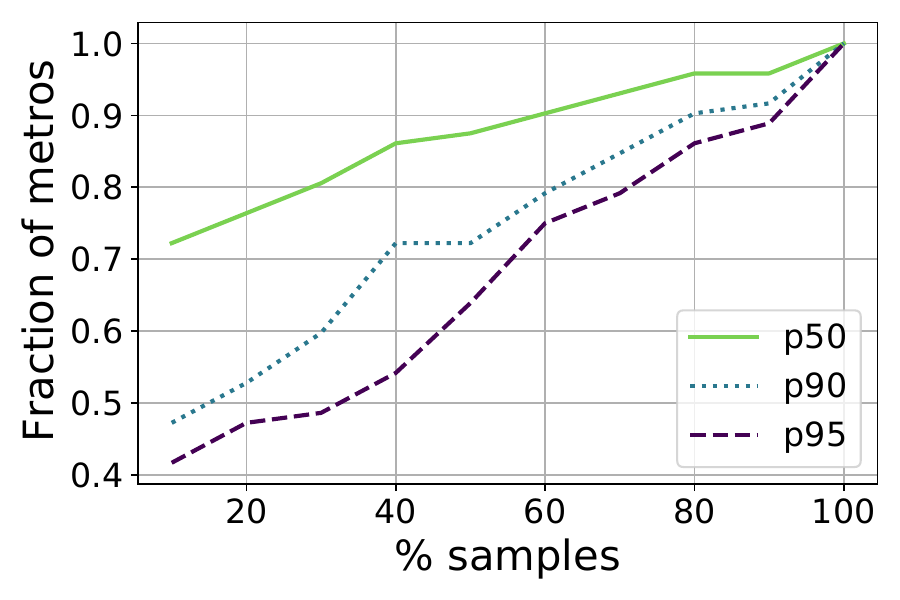}
             \caption{Faster PoP \%  (24 hours).}
             \label{fig:non_default_count_24_hours}
         \end{subfigure}
          {\vspace{0.15in}}

          \caption{
            \ref{fig:non_default_count} shows \% of samples a non-default PoP performed better 
          than geo-default. \ref{fig:anycast_non_default_count} shows the number of times a faster non-default PoP was the same as the
          anycast-default for metros where the geo-default was different from the anycast-default. \ref{fig:non_default_count_24_hours} 
          shows performance of non-default PoPs remains stable for a long time.}
          \label{fig:latency_analysis}
          {\vspace{0.1in}}
\end{figure*}

Table~\ref{tab:analysis_data} summarizes the data we use. Since our goal 
is to identify better alternatives for placing 
a WAN hub, we focus on metros with measurements 
to more than one PoP.

\subsection{Quantifying the opportunity}
\label{sec:data_analysis}

We analyze the ratio between minimum latency to the geo-default PoP
(current practice \S\ref{sec:background}) and other PoPs
to quantify the opportunity of improving connection latencies
by placing hubs in alternate locations.
Minimum latency is less affected by
transient phenomenon (\eg congestion, queuing delays) on measured latency.
We find the minimum latency of the geo-default PoP and  
all other PoPs in a ten-minute time window for each client metro. 
Some metros have more uniform measurements over time and contribute 
more samples for comparison in more ten-minute time windows than others.
To prevent these metros from skewing our analysis, we
normalize the sample sizes by selecting 50 random samples across
all windows for each client metro. We discard metros with 
fewer than 50 samples. 

Figure~\ref{fig:latency_ratio} shows the cumulative distribution
function (CDF) of log$_{2}$ ratio of minimum latency of the geo-default PoP
to the fastest non-default PoP. A point above 2 on the x-axis
  indicates a 4x latency improvement. The graph shows the median, 90$^{th}$ and
95$^{th}$ percentile latency computed over ten-minute time windows for
100 metros (5000 data points). 
For $\approx$20\% of the windows,
placing a hub at a PoP that is not geographically closest
would result in better median latency for the enterprise. The
performance gap between geo-default and other PoPs is wider at higher
percentiles.

\myparab{Are alternate PoPs faster than geo-default?} The
previous results were across all sampled time windows for all
metros. We now turn to single time windows.  From
our 50 windows for each metro, Figure~\ref{fig:non_default_count}
shows the number of sampled windows where the fastest PoP was not the
geo-default PoP. For $\approx$75\% of metros, there were only
0-5 samples out of 50 where any PoP had lower median latency than the
geo-default. However, for 90$^{th}$ and 95$^{th}$ percentiles, this drops to 58\%
and 47\% of metros respectively, indicating tail latencies are more often better
at other PoPs. Rarely was another PoP always fastest
(around 100\% samples), but there is a fairly even distribution of how
often an alternate PoP was faster than the geo-default.
 \emph{There is an opportunity to
  improve latency by placing hubs using a
  dynamic mechanism.}

\myparab{How often are faster PoPs the same as anycast-default?} 
For 20 out of 100 metros,
geo-default and anycast-default PoPs were different. 
Figure~\ref{fig:anycast_non_default_count} shows, for the 20 metros, how
many samples found the anycast-default PoP the fastest at different
percentiles. 
The results show that for
8 metros (40\%), the anycast-default had lowest median latency in only 0--5 of
50 samples, and only 3 metros (15\%) found that anycast-default was almost
always the fastest.
 \emph{Anycast often does not always select the lowest-latency PoP, so
   there is an opportunity to improve latency by placing a hub at a location which is not chosen often by anycast.}

\myparab{Sustained performance benefit from faster PoPs.} Our final
question is how long can an alternate PoP sustain its performance
benefit; it may be that performance gains are transient, so the delay
of reconfiguring a network precludes using a faster PoP.
We pick 50 samples in a large window: 1 hour, 10
hours, or 24 hours. 
We count how many times an alternate PoP had lower latency (at a percentile) than the
geo-default PoP. We identified 95 metros with
enough data for samples at 1-hour granularity, 84 at 10 hours, and 72
at 24 hours. Figure~\ref{fig:non_default_count_24_hours} shows the
results for 24-hour windows; results for 1 hour and 10 hours show the
same trends (See~\ref{sec:appendix_non_default_count}). This graph shows the 
number of metros that had access to faster alternate PoPs (x-axis). 
For larger windows, it is {\em even more likely} that an
alternate PoP will outperform the geo-default over 24
hours. 
 \emph{Thus, faster PoPs perform better than geo-default for long periods 
 and optimizing the latency of these metros can enable sustained performance improvements.}




\subsection{Quality of decision-making}
\label{ref:quality_dataset}
\begin{wrapfigure}{R}{0.45\textwidth}
  \centering
  \includegraphics[width=0.45\textwidth]{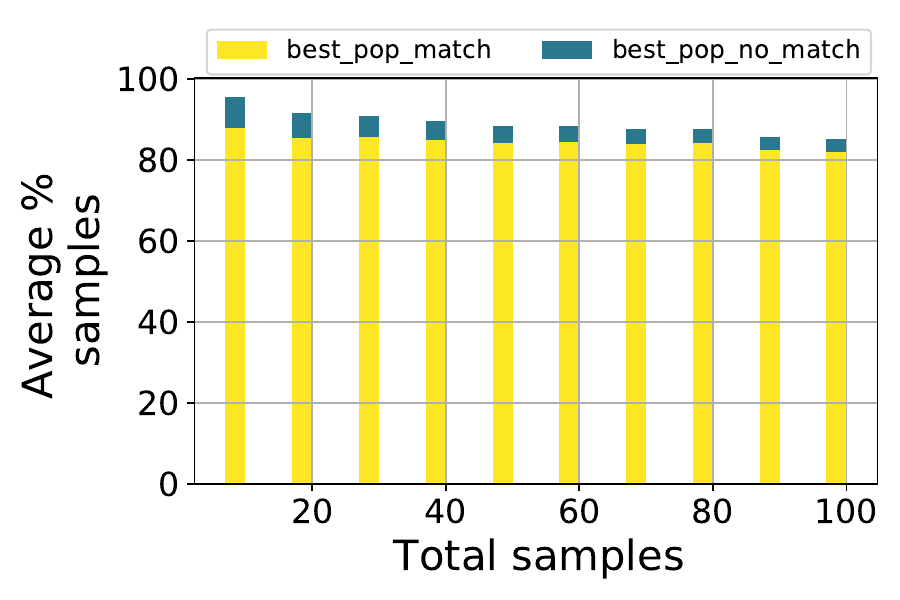}
  \caption{PoP 
  matches or no matches for the \emph{best-pop} over selected sample size with \emph{best-pop-overall}. 20 samples
  are enough to make an accurate decision $\approx$75-85\% of the times.}
  \label{fig:policy_analysis_90}
\end{wrapfigure}

Real-world datasets may not have measurements for all
metros for all time periods; there may sometimes be only
a few measurements available. 
To understand how having limited data impacts decision-making, 
we artificially reduce the sample size of measurements and compare
decisions made with ample data \vs those made with smaller samples.
Using samples from a 2-day
window, we compare the lowest-latency PoP identified by averaging a
small number of samples (10, 20, 30 etc.), labeled {\em
best-pop}, against the lowest-latency PoP considering all measured
data, labeled {\em best-pop-overall}. We run this experiment 10 times 
to test on different sets of random samples of each size, and report
the average percentage of matches over all runs. 
Figure~\ref{fig:policy_analysis_90} shows how often {\em best-pop}
matches {\em best-pop-overall} and how often it is different
(no-match). Not all windows have the required number of samples, so
the two groups do not add to 100\%. These results show that with only
20 measurement samples per window, we can make correct decisions about
$\approx$85\% of the time, and that adding more samples does not help.
\emph{We conclude that averaging latencies from 20 samples 
identifies the fastest PoP with a probability of 80\%.}

\section{Evaluation}
\label{sec:evaluation}

\sysname aims to produce better virtual WAN topologies than
the default choice of cloud PoPs for enterprise branch offices. 
We select the geographically closest PoP as the default 
({\em geo}) and also compare against the anycast-default ({\em anycast}).
We evaluate \sysname's ability along two axes:
(1) Can \sysname improve the connection latency of a virtual WAN?
and (2) Can \sysname reduce the total hubs needed by an enterprise? 
We also evaluate \sysname's ability to balance these two goals.

\subsection{Methods}
\label{sec:methods}
We implement \sysname's optimization algorithm
in Python 3 using the convex optimization language 
CVXPY~\cite{ref:cvxpy_1,ref:cvxpy_2}. CVXPY solves our mixed integer
linear problem (MILP) formulation efficiently by choosing a solver
suitable for the objective (Alg.~\ref{alg:cap}).

For all formulated topologies in our evaluation, the solver takes at
most $\approx$ 19 \emph{sec}.  In (\S\ref{sec:maintain}), we show that
reconfiguring topologies every 4 days can further improve
performance. This cost, even if the optimization slows down for larger
topologies, is negligible as customers only need to run it every 4 days.

\begin{wraptable}{R}{0.42\textwidth}
  \small
  \centering
    \begin{tabular}{|l|r|}
    \hline
    Total samples      & 3,531,541                      \\ \hline
    \begin{tabular}[c]{@{}l@{}}Total samples of metros \textgreater 1
      PoP\end{tabular} & \multicolumn{1}{l|}{2,530,985} \\ \hline
      Total PoPs & 17 \\ \hline
      Total metros & 281 \\
      \ \ \ \ Metros with 1 PoP & 152   \\    
    \ \ \ \ Metros \textgreater 1 PoP(s)    & 129  \\ \hline
    \end{tabular}
    \caption{Evaluation data summary}
    \label{tab:evaluation_data}
\end{wraptable}

We evaluate the optimization algorithm on the latency measurement dataset
collected in Oct --- Nov 2021. Table~\ref{tab:evaluation_data} summarizes the
data. We only include metros having at least 20 measurements each to more than one 
PoP. We choose this sample size based on the
data in Figure~\ref{fig:policy_analysis_90} showing that 20 samples is
sufficient. After this filtering, 
95 metros remain.
We measure the overall latency of a
topology as the {\em average latency across all connections}: we
calculate the average latency across a customer's metro regions
weighted by the number of connections from the region. 
As stated in
\S\ref{sec:hub_placement}, we set the connection limit per hub to
1000 (\eg a metro with 3200 total connections will need minimum of 4 hubs).  

\myparab{Synthetic customer topology.}
We generate {\em synthetic
client topologies} of sizes 5, 10, 25, 50, and 75 branch office locations
(metros).  We chose these topologies sizes because 
they realistically represent companies of small and large sizes 
(\eg a large company like Puma~\cite{ref:puma-loc} has $\approx$35 total office locations worldwide).
The metro locations for each are randomly 
selected from the set of metros. For each size, we generate 100 topologies (500 in total).
For total connections from a
branch office, we sample a number from an exponential distribution
with a 90\% probability of the number being between 100 and 10,000. The largest generated topology has 
93052, and the smallest topology has 1019 total connections. The synthetic topology
generation does not cover all the possible company and
organization topologies, such as a
retail company with a large headquarter and many similar-sized branch offices. However, we think that these synthetic topologies cover an adequate diversity of different
sizes (small, medium, and large) to show the potential improvements in latency and reductions in cost with our proposed method.

\begin{wraptable}{r}{0.38\textwidth}
  \small
    \begin{tabular}{|r|r|r|r|}
        \hline
        metro     & clients & default PoP & lat\\ \hline 
        Louisiana & 825     & dal         & 31   \\ \hline 
    \end{tabular}
    \caption{Partial enterprise topology.}

    \label{tab:example_topo}
\end{wraptable}

\myparab{Provider topologies.}  The hub placement algorithm takes
synthetic topologies as input (\S\ref{sec:hub_placement}). We
simplify the inputs to include PoPs for which at least one
customer metro has latency measurements. We calculate the latency 
values from each metro to each PoP at the 90$^{th}$ percentile 
threshold to populate ${L}$ in Alg~\ref{alg:cap}. 
PoPs for which there is no data are given
a very high latency to prevent them from being
selected. Table~\ref{tab:example_topo} 
shows an example enterprise topology: 
the customer has a branch office in Louisiana with 825 client
connections and 
a default PoP of Dallas (dal), with measured latency of at most 31\emph{ms}
for 90\% of the time in the interval selected. We calculate the overall latency of
a topology as described above.

\subsection{Latency vs Cost Tradeoffs}
\label{sec:eval}
We evaluate the latency benefits for
\stt{l\_optimal}, and the performance inflation (if any) for \stt{k\_optimal}.
We use hub count as a proxy to measure cost for both modes.

\begin{figure*}[t]
   \hfill
  \begin{subfigure}[t]{0.44\columnwidth}
      \centering
      \includegraphics[width=\columnwidth]{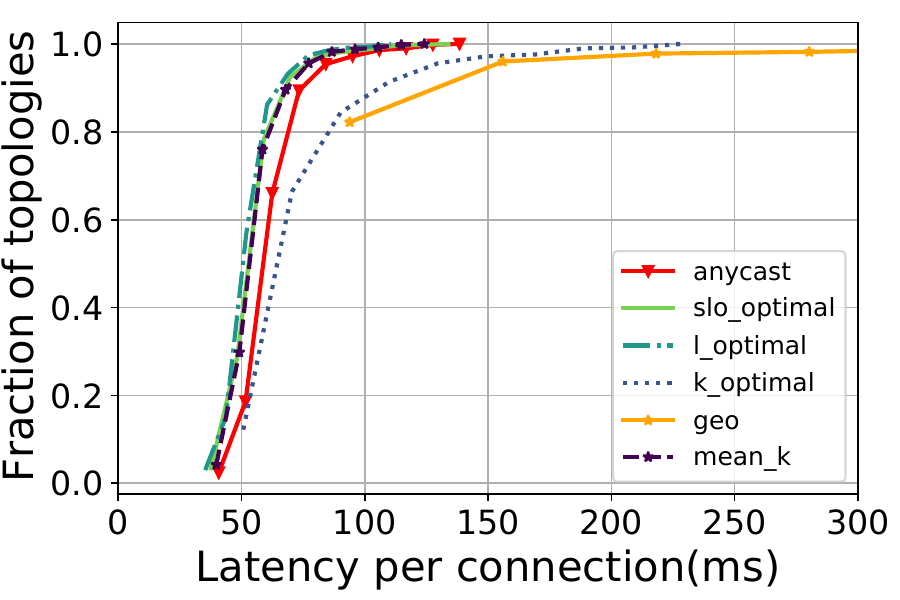}
      \caption{CDF of latency per connection.}
      \label{fig:lat_90}
  \end{subfigure}
   \hfill
  \begin{subfigure}[t]{0.44\columnwidth}
    \includegraphics[width=\columnwidth]{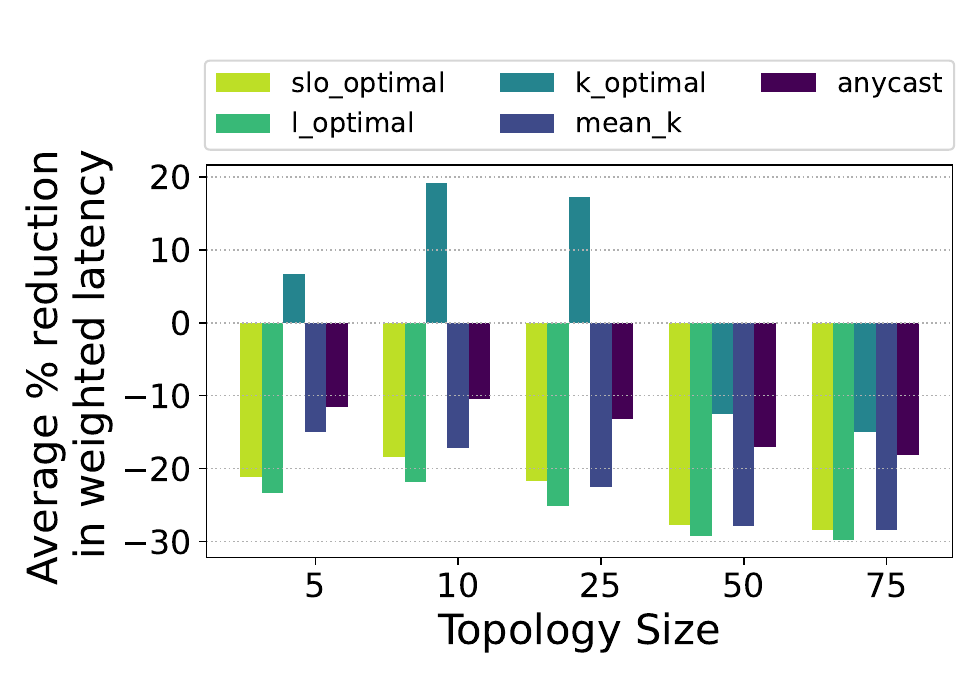}
     \caption{\emph{geo} latency reductions.}
     \label{fig:lat_red_90_geo}
\end{subfigure} 
  \hfill

  {\vspace{0.12in}}
  \caption{~\ref{fig:lat_90} shows CDF of weighted connection
    latency for different policies. We set the max to
    300\emph{ms} on the \emph{x-axis} to remove long tail of
    \emph{geo} for better readability. ~\ref{fig:lat_red_90_geo} shows the reduction in weighted
    latency with \emph{geo}.}
    {\vspace{0.1in}}
\end{figure*}

We compare the latency per connection in topologies generated by \sysname
with baselines.
Figure~\ref{fig:lat_90} shows the CDF of latency per connection across all 500 topologies
(full CDF in Appendix~\ref{sec:appendix_lat}). We observe that the
\emph{anycast} strategy performs much better than \emph{geo}. For 
$\approx$97\% of the topologies the latency per connection is less than
$\approx$80\emph{ms}, and for all topologies it is less than $\approx$125\emph{ms}
in the \stt{l\_optimal} mode. This mode outperforms \emph{geo} and \emph{anycast} 
for all topologies. \stt{k\_optimal} performs worse than \emph{anycast}, 
but does better than \emph{geo} for some metros. 

Table~\ref{tab:lat_reduction_both_baselines} 
summarizes average latency reductions. Overall, \stt{l\_optimal} 
reduces latency by 26\% compared with the \emph{geo} baseline.
\emph{anycast} is 14\% faster than \emph{geo} and
13\% slower than \stt{l\_optimal}. 
Observe in Figure~\ref{fig:lat_red_90_geo} that latency reductions are greater for larger topologies,
where there are more options to avoid bad hubs, and a
greater likelihood of having a branch office in a metro where the
closest PoP performs poorly (so \emph{geo} does poorly). For smaller
topologies, there may be several nodes with few good alternatives. 
\stt{k\_optimal} has higher latency than \emph{anycast} for all
topology sizes, but the added latency is low for larger topologies.
Compared to \emph{geo}, \stt{k\_optimal} has only 3\% higher latency 
on average across all topologies, and substantial reductions for larger 
topologies, where there are more optimization opportunities.
This shows that making hub placement decisions using \sysname
can greatly reduce connection latency.


\begin{wraptable}{R}{.57\textwidth}
  \small
  \begin{tabular}{|r|r|r|r|r|r|}
  \hline
          & l\_optimal & k\_optimal & mean\_k & slo  & anycast \\ \hline
  geo     & -26\%       & 3.1\%        & -22\%    & -23\% & -14\% \\ \hline
  \end{tabular}
 
  \caption{Average weighted latencies reduction by \sysname.}
  \label{tab:lat_reduction_both_baselines}
\end{wraptable}

\begin{figure*}[t]
  \centering
  \begin{subfigure}[t]{0.32\textwidth}
    \centering
    \includegraphics[width=\columnwidth]{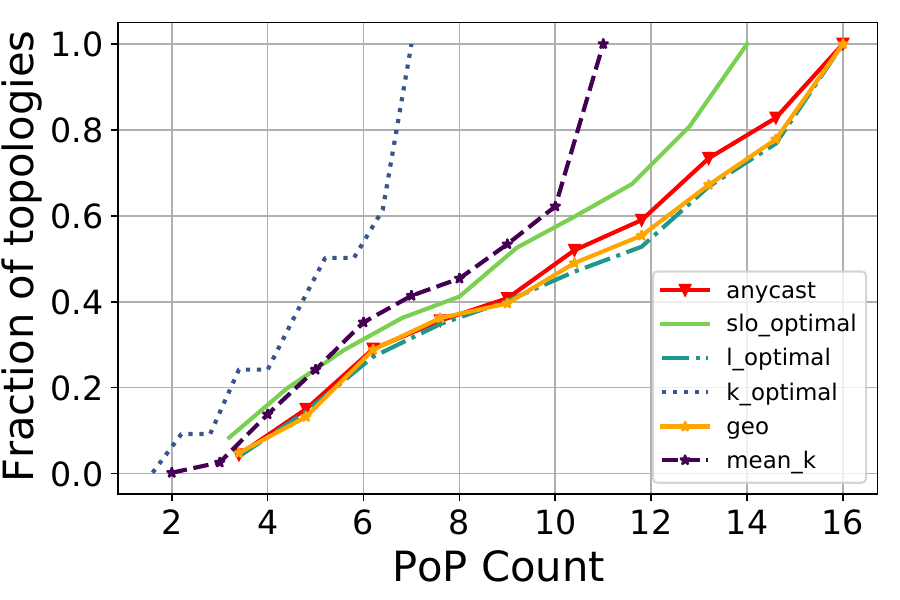}
    \caption{CDF of PoP counts.}
    \label{fig:pop_size_90}
  \end{subfigure}
  \hfill
  \begin{subfigure}[t]{0.32\textwidth}
    \centering
    \includegraphics[width=\columnwidth]{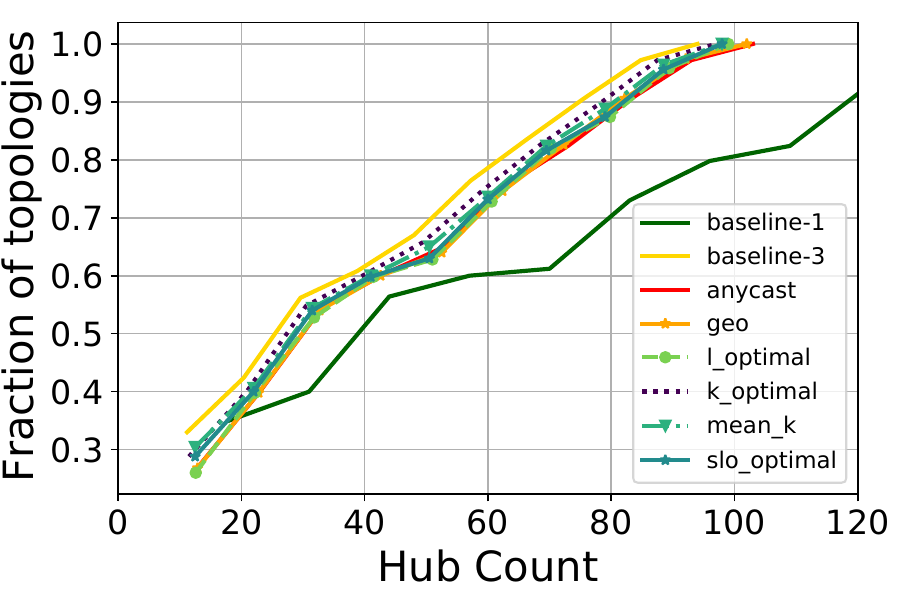}
    \caption{CDF of hub counts.}
    \label{fig:hub_size_90}
  \end{subfigure}
  \hfill
  \begin{subfigure}[t]{0.32\textwidth}
    \centering
    \includegraphics[width=\columnwidth]{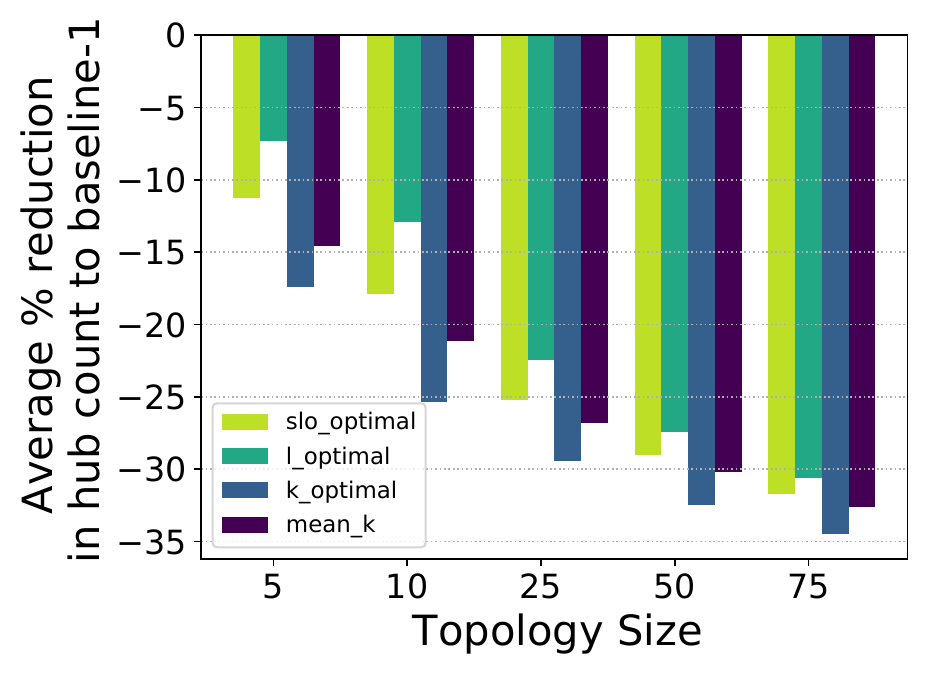}
    \caption{Reduction in hub counts. }
    \label{fig:hub_size_red_90}
  \end{subfigure} 

  {\vspace{0.15in}}

  \caption{~\ref{fig:pop_size_90} shows the total number of PoPs used in each mode with the least being in \stt{k\_optimal}.
   ~\ref{fig:hub_size_90}-~\ref{fig:hub_size_red_90} show the impact of the PoPs used on the hub counts in comparison to all baseline. 
   All modes outperform \emph{baseline-1} with \stt{k\_optimal} with the most reduction of 28\%, followed by 
   \stt{slo\_optimal}(23\%) and \stt{l\_optimal}(20\%).  \emph{geo} is very close to \emph{anycast}.
   \stt{k\_optimal} comes closest to \emph{baseline-3}, \stt{l\_optimal} performs 
   no worse than \emph{geo}, and \stt{slo\_optimal} does
   slightly better than \emph{geo}.}
   {\vspace{0.1in}}
\end{figure*}

\myparab{Topology Sizes.}
There is a fixed cost of
provisioning a hub (\eg a standard hub on Azure costs upto \$3500/year~\cite{ref:vwanpricing}), so we use the total hubs needed for a
topology as a proxy for cost. Figure~\ref{fig:pop_size_90} shows 
the total unique PoPs used by both modes. This is important as the total hubs
depend on the PoP counts as stated (\S\ref{sec:hub_placement}). We
observe that \stt{k\_optimal} uses the fewest unique PoPs and is 56\% lower than the baselines for all the topologies.
\stt{l\_optimal} uses more PoPs than baselines for all topologies.

To understand the impact of PoP count on the number of hubs, we
compare \sysname's \emph{hub} placement  against three baselines:
(1) For \emph{baseline-1} each metro has its own set of
  hubs. We first divide the number of connections per
  metro by ${\beta}$, and then take the sum over all metros. This
  approach is commonly used today by enterprise customers. 
(2) For \emph{baseline-2} metros share hubs. We calculate
  the number of connections to each PoP from all metros, and divide by
  ${\beta}$ to get the number of hubs at the PoP, and then sum across
  all PoPs.
(3) For \emph{baseline-3} we calculate the minimum hubs
  irrespective of topology. We sum the required connections from each
  metro and divide by ${\beta}$ to get the total number of
  hubs. This is theoretically the lowest number of hubs possible.
\emph{baseline-1}, and \emph{baseline-3} are same for both \emph{anycast}, 
and \emph{geo}. We calculate the number of hubs for \sysname,
\emph{anycast} and \emph{geo} using \emph{baseline-2}.

We show the CDF of total hubs for all baselines and \sysname's different modes in
Figure~\ref{fig:hub_size_90}. First, we observe the importance of sharing
hubs across client metros: \emph{baseline-1} does significantly worse
on all topologies across all modes. Second, \emph{baseline-2}
for \emph{geo} and \emph{anycast} are very close. This is because
\emph{geo} or \emph{anycast} assign hubs independently for each
metro without attempting to share hubs. Third,
\stt{k\_optimal} outperforms both \emph{baseline-2}, and comes very close to \emph{baseline-3}. Finally, \stt{l\_optimal} 
is just as good as both \emph{baseline-2} (\emph{geo} and \emph{anycast}). This implies that even with the 
same cost, we can achieve better performance by placing the PoPs differently.

\begin{wraptable}{R}{.54\linewidth}
  \small
  \begin{tabular}{|r|r|r|r|r|}
  \hline
          & l\_optimal & k\_optimal & mean\_k & slo    \\ \hline
  baseline-1 & -20\% & -28\%     & -25\%  & -23\% \\ \hline
  baseline-2 & 1.15\% & -8.39\%     & -4.92\%  & -2.47\% \\ \hline
  \end{tabular}
 
  \caption{Average hub count reduction with \emph{baseline-1} and \emph{baseline-2} (anycast).}
  \label{tab:avg_hub_reductions}
\end{wraptable} 

We compare the reduction in hub counts for different topology sizes with \emph{baseline-1} shown in Figure~\ref{fig:hub_size_red_90}.
Average hub reductions across all topologies is shown in Table~\ref{tab:avg_hub_reductions}.
These show that \stt{k\_optimal} has $\approx$28\% fewer hubs on average across all
topologies. \stt{l\_optimal}, despite being the most performant, also
reduces hub count by an average of $\approx$20\%. Most of this benefit
comes from sharing hubs across branch offices. Reduction in hub count with
\emph{baseline-2} is minor for all modes except \stt{k\_optimal}, which uses $\approx$8\% fewer hubs.
These results indicate that \sysname can minimize the total hubs
saving thousands of dollars in cost. 

\myparab{Summary.}
\label{sec:eval_insights}
There is a benefit in both performance and
cost when hubs are placed differently.
The \stt{l\_optimal} has on average 26\% lower
latency than {\em geo},
and \stt{k\_optimal} uses 28\% fewer hubs.

\subsection{Policy choices}
\label{sec:policy_choices}
\begin{figure*}[t]
  \centering
  \begin{subfigure}[t]{0.32\textwidth}
    \centering
    \includegraphics[width=\columnwidth]{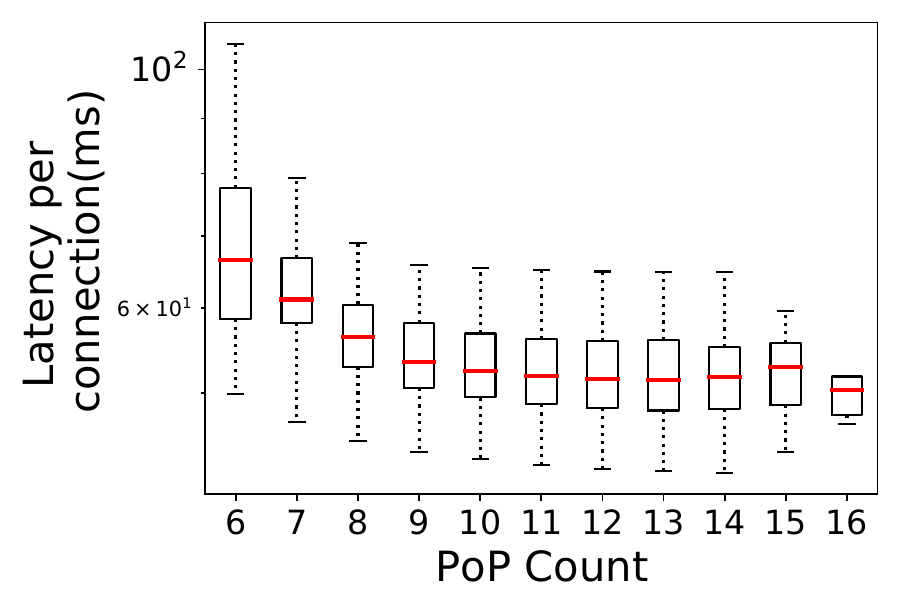}
    \caption{Distribution of latency per connection for Topology of size 50.}

    \label{fig:pareto_50}
  \end{subfigure}
  \hfill
  \begin{subfigure}[t]{0.32\textwidth}
    \centering
    \includegraphics[width=\columnwidth]{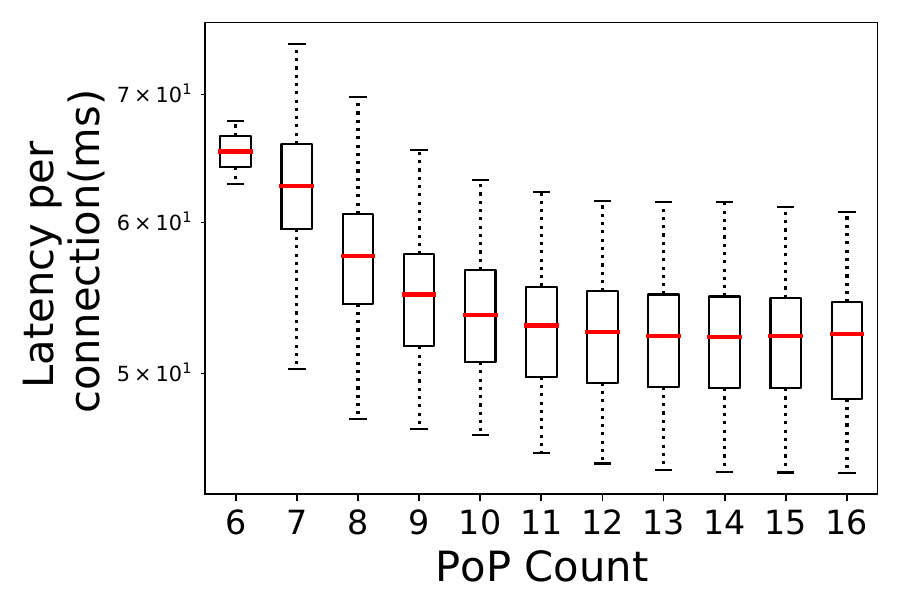}
    \caption{Distribution of latency per connection for Topology of size 75.}
    \label{fig:pareto_75}
  \end{subfigure} 
  \hfill
  \begin{subfigure}[t]{0.32\textwidth}
    \centering
    \includegraphics[width=\columnwidth]{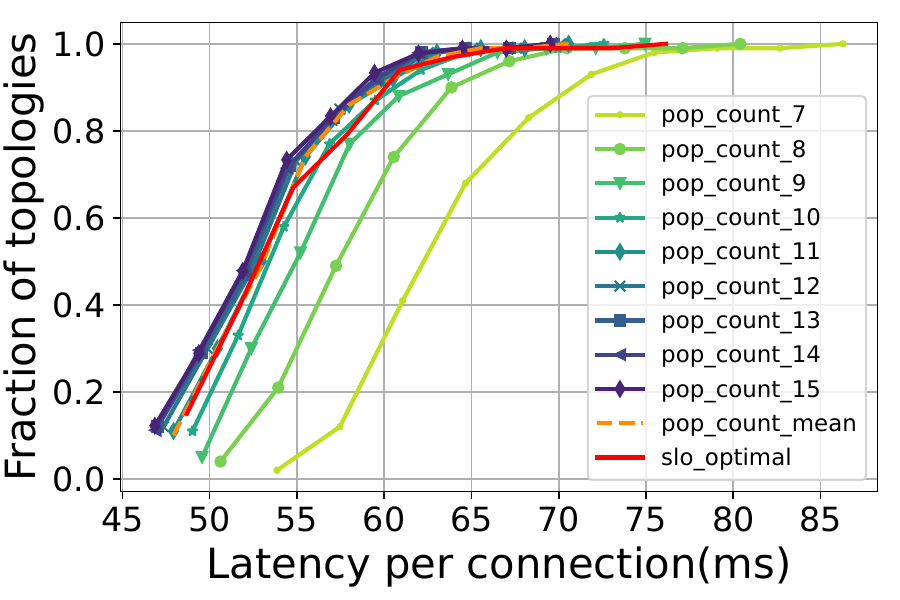}
    \caption{Latency at each PoP count for topologies of size 75.}
    \label{fig:all_pop_lat_75}
    
  \end{subfigure} 

   {\vspace{0.15in}}
  \caption{~\ref{fig:pareto_50}-~\ref{fig:pareto_75} show that regardless of the topology size, 
  pop count of \stt{mean\_k} bounds the latency benefits. ~\ref{fig:all_pop_lat_75} shows 
  latency per connection at each PoP count for topology size 75. \stt{mean\_k} performs significantly 
  better with little additional cost. Thus, \stt{mean\_k} is a good heuristic to optimize for both, 
  performance and cost.}
  {\vspace{0.1in}}

\end{figure*}

The results of \stt{k\_optimal}, and \stt{l\_optimal} sit on the opposite ends of the 
cost vs. performance Pareto frontier.

This is problematic when
customers want to balance performance and cost. A good solution 
balances the value of ${K}$ to 
save cost while being close to \stt{l\_optimal}. 
To get this solution, we need to evaluate all solutions on the
Pareto frontier -- topologies with minimum latency for a given $K$ such that
reducing K increases latency.

\myparab{Pareto frontier.} We explore all possible solutions on the Pareto Frontier.
We do this by minimizing the latency at each possible $K$ value between $k_{min}$ and $k_{max}$ for each topology. 
We show the box plots of the distribution of 
latency per connection at each PoP counts for 100 topologies of sizes 50 and 75 in Figure~\ref{fig:pareto_50}-~\ref{fig:pareto_75}. Other topology 
sizes show similar trends (Appendix~\ref{sec:appendix_pareto}). We include all PoP counts in the 
Figure~\ref{fig:pareto_50}-~\ref{fig:pareto_75} where at least 1
topology gave a feasible solution.

As we increase the number of PoPs, the reduction in latency diminishes 
after around the mean value of $k_{min}$ and $k_{max}$ (Appendix~\ref{sec:appendix_pareto_curve}). The variance at $k_{min}$ and
$k_{max}$ occur because some topologies do not have solutions for small
and large $K$. 
This result shows there is an initial rapid reduction in latency as
PoPs are added to a topology, but at the midpoint between $k_{min}$ and
$k_{max}$ the curve flattens and adding more PoPs has little latency benefit. 

\myparab{Mean-K}. Based on our observations of the Pareto frontier, we
propose a heuristic policy called \stt{mean\_k} that chooses the
latency-optimal topology with the number of PoPs equal to the mean value
of $k_{min}$ and $k_{max}$.  We show that this
heuristic can optimize performance with a high probability without
incurring much cost, which is helpful for customers wanting both low
cost and low latency. To verify our observation, we
plot the latency for \stt{mean\_k} in Figure~\ref{fig:lat_90}. 
We observe that \stt{mean\_k} is close to \stt{l\_optimal}, and
for all topologies it outperforms both baselines.
Figure~\ref{fig:all_pop_lat_75} shows the CDF of latency per
connection at each PoP counts for topology size of 75.  Other sizes
show a similar pattern (Appendix~\ref{sec:appendix_lat_dist}).  These
show that by using \stt{mean\_k}, we can have significant performance
benefits with little additional cost. We also show the reduction in
weighted latency in
Figure~\ref{fig:lat_red_90_geo} against \emph{geo}. 
Table~\ref{tab:lat_reduction_both_baselines} shows the 
average across topologies. The latency 
improvement for \stt{mean\_k} is lower
than \stt{l\_optimal} by 4 percentage points. It reduces the number
of hubs by 25\%, making it only 3 percentage points more expensive
than \stt{k\_optimal} on average (Figure~\ref{fig:hub_size_red_90} and
Table~\ref{tab:avg_hub_reductions}).

\myparab{Summary.}
\label{sec:pareto_insights}
The Pareto frontier allows a customer to trade off performance and
cost. The \stt{mean\_k} policy balances both goals to achieve
near-optimal latency and near-optimal cost.

\subsection{Optimization extensions with SLOs}
Our optimization can be extended to include service-level objective (SLO) that
keeps latency within some threshold, such as some branch offices
achieving 50\emph{ms} and others 100\emph{ms}. 

\myparab{Changes in the algorithm.}The optimization needs the latency 
requirement input from the customer.
This can be captured in the input variable, $\alpha_{j}$: desired latency for connections to metro $j$.
We also add a new demand constraint:
\begin{equation} \tag{6}
  \sum_{i=1}^{p} U_{ji} \cdot L_{ij} \leq \alpha_{j},   \forall j \in m
    \label{ref:const_lat}
\end{equation}

This ensures that any metro/PoP combination chosen by the algorithm
(where $U_{ji} = 1$) will have latency below $\alpha_j$. 
The objective function minimizes the latency with this additional constraint. We introduce a new mode, \stt{slo\_optimal} to 
solve the optimization with these changes. 

\myparab{Evaluation.}A realistic SLO was not available to us so we use
\emph{anycast} latency as the default SLO for our evaluation. 
This ensures that the new topologies are not any worse than what anycast chooses. 
\stt{slo\_optimal} optimizes both latency and hub counts while meeting the latency cap per metro.
We show the CDF of latency per connection with \stt{slo\_optimal} in Figure~\ref{fig:lat_90}. We observe that all topologies
outperforms \emph{anycast}
with an average reduction of $\approx$11\% in weighted latency shown in 
Table~\ref{tab:lat_reduction_both_baselines}. 
Figure~\ref{fig:pop_size_90} shows total PoPs used and Figure~\ref{fig:hub_size_90} shows the hub counts.
The average reduction in hub count is 23\% (Table~\ref{tab:avg_hub_reductions}).
This shows that the same PoP can meet the SLO for multiple metros, 
which may not always be the \emph{anycast} or \emph{geo} PoP.

\section{Maintaining Virtual WANs}
\label{sec:maintain}

Our evaluation has shown the benefits of optimized
virtual WAN hub placements when applying
\sysname retroactively, i.e., over historic data. In this section we
show how well \sysname works when used {\em predictively}: making
placements based on recent measurements, and evaluating the subsequent improved
performance.
We evaluate how to {\em maintain} a virtual WAN topology with three
questions: (1) Can \sysname make good predictions of future latency?
(2) How much historical data is needed for placement decisions? (3) How often do
\sysname generated topologies need to be updated?

To evaluate these questions, we measure the impact of how data 
is {\em smoothed}  and how often topologies should be regenerated. Using an exponentially weighted 
moving average, we evaluate how it should be weighted to balance historical
data against new data. We look at values from 1 (no historical data) to 0.1 (slow moving, mostly
historical data). With this smoothing, we run the optimization on
different window sizes, reconfiguring every day, 2 days, and 4 days. 
To ensure that each metro 
in the topology under test has enough
samples, we start with the data used for the evaluation period
(October-November 2021) of 45 days and break the observation data into
non-overlapping 1-day, 2-days and 4-days windows. We filter out metros
with fewer than 20 samples, leaving 59 metros.  We create 10
topologies each of sizes, 5, 10, 25, 35, and 45 (50 total).
Because this is a different set of metros, the results in
this section are not directly comparable to those in
\S\ref{sec:evaluation}. Instead, we include results running the same
experiments on the reduced set of metros.

We calculate the 90$^{th}$ percentile latency for metro 
$j$ in connecting to PoP $i$ in time window $w$ as $\alpha_{ij,w}$. 
To smooth data for optimization,
we calculate the input $L_{ij,w}$ (latency from metro $j$ to PoP $i$
in time window $w$) by varying the amount of smoothing, $\gamma$ as:
\begin{equation} \tag{7}
    L_{ij,w} = \gamma \cdot \alpha_{ij,w} + \left( 1 - \gamma \right) \cdot  \alpha_{ij,w-1}
    \label{eq:moving-avg}
\end{equation}

\begin{wrapfigure}{R}{0.4\textwidth}
  \centering
  \begin{subfigure}{0.4\columnwidth}
    \includegraphics[width=\columnwidth]{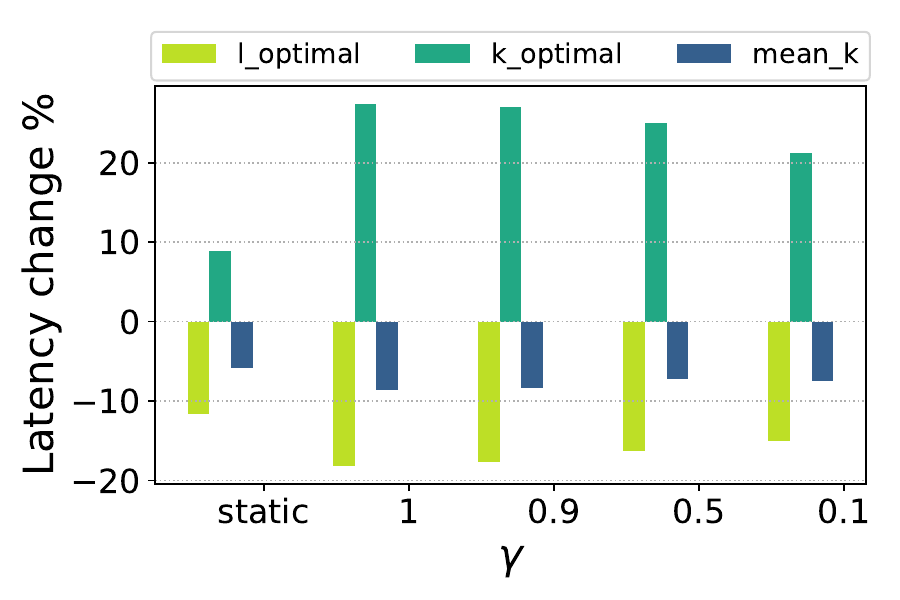}
    \caption{Latency changes.}
    {\vspace{0.05in}}
    \label{fig:lat_maintain_with_geo_4_days}
  \end{subfigure}
  \begin{subfigure}{0.4\columnwidth}
    \includegraphics[width=\columnwidth]{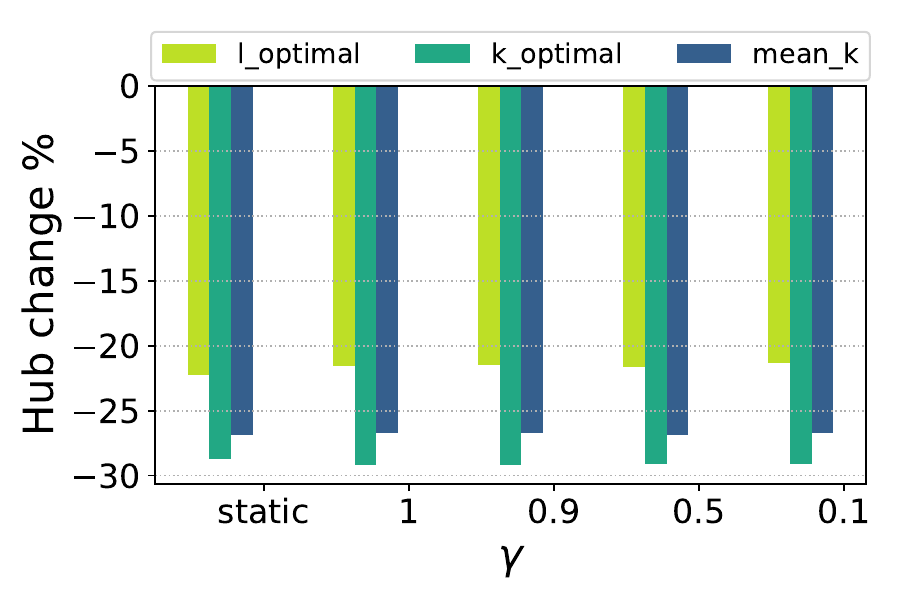}
    \caption{Hub count changes.}
    \label{fig:hubs_maintain_with_geo_4_days}
   \end{subfigure} 

   {\vspace{0.15in}}
  \caption{~\ref{fig:lat_maintain_with_geo_4_days}-~\ref{fig:hubs_maintain_with_geo_4_days} shows the latency, and hub with different 
    smoothing variations with geo baseline on 4-days window.}
\end{wrapfigure}

\myparab{Results.} We present the results when updating topology
every 4 days for latency for the
\stt{l\_optimal}, \stt{k\_optimal}, and \stt{mean\_k} policies
in Figure~\ref{fig:lat_maintain_with_geo_4_days},
and hub count in Figure~\ref{fig:hubs_maintain_with_geo_4_days}. 
All these results are averaged across all topologies of all sizes. 
Both graphs show the average percent change in latency and hub count over all
windows relative to the geographical placements over 45 days. We also show the 
placements from \S\ref{sec:eval} using latency data averaged over the
whole 45 days (marked as \emph{static}). The \emph{static} latency
reductions here are lower than in Table~\ref{tab:lat_reduction_both_baselines} 
because the topologies for \emph{static} use a subset of metro-PoP combinations 
(59 metros vs 129 in \S\ref{sec:eval}). This limits and makes latency choices 
different than \S\ref{sec:eval}.

The different smoothing variations perform better than \emph{static} for \stt{l\_optimal} and 
\stt{mean\_k} (Figure~\ref{fig:lat_maintain_with_geo_4_days}). All modes
show similar reductions in hub count compared to \emph{static} (Figure~\ref{fig:hubs_maintain_with_geo_4_days}).
These results indicate that adapting predictively is better for
latency and does not affect cost. 

Looking now at the impact of smoothing,
for \stt{l\_optimal}, Figure~\ref{fig:lat_maintain_with_geo_4_days}
shows that smoothing is not
necessary for latency, and it yields lowest latency 
with no ($\gamma$ = 1) or little historical data ($\gamma$ = 0.9). For \stt{k\_optimal},
using slow moving average ($\gamma$ = 0.9) performs the best.
\stt{mean\_k} follows the pattern of \stt{l\_optimal}, and performs best with less smoothing (fast moving average).
The results for 1-day and 2-day windows are almost identical (Appendix~\ref{sec:appendix_maintain}). 
These results show that placements are stable for extended periods, and do not need to be updated frequently. However, 
re-computing placements with little to no historical data, and updated measurements 
every 4-days show up to $\approx$10\% latency improvements. These
indicate that we need some maintenance to keep performance high but not much.

Turning to the cost (number of hubs change with \emph{baseline-1}) in Figure~\ref{fig:hubs_maintain_with_geo_4_days},
for all the modes the changes remain insensitive to the amount of smoothing and are identical to \emph{static}.
The total reductions across all the modes also remain stable for different windows (Appendix~\ref{sec:appendix_maintain}).
These results show that there is no affect on cost by reconfiguring
topologies every few days, further showing that \sysname is able to
make placement choices that continue to give benefits over a period
into the future. 

\myparab{Summary.} \sysname-generated topologies remain stable for extended periods. 
Customers can choose to reconfigure topologies every 4-days to get
peak latency but needs no configuration for cost. Furthermore, a key benefit of our approach 
is that \sysname's optimization formulation can be extended to include other constraints,
such as a cap on the number of hubs that are moved during reconfiguration.
This is a straightforward extension to
limit the magnitude of difference between old and new topologies.

\section{Discussion}
\label{sec:discuss}
In this section, we identify the vehicle for \sysname's adoption by
enterprises or cloud providers. We then discuss alternative mechanisms
that can resolve performance anomalies observed by enterprise virtual
WANs today.

\myparab{Adoption of \sysname by cloud networks.}
While \sysname can operate as a standalone tool for enterprises looking
to design their virtual WAN deployments, we believe it can be 
integrated by cloud providers into their portals.
For instance, \sysname can be integrated into Microsoft's Azure portal~\cite{ref:azportal}
to suggest enterprises how many virtual WANs should they spawn based in their optimization goal(s). 
This is similar to suggestions made by cloud providers
when users spawn virtual machines today.

\myparab{Interventions from cloud providers.}
An alternative to enterprise customers improving their
virtual WAN performance through topology design is
intervention by cloud providers when performance anomalies occur.
The cloud provider in Figure~\ref{fig:vwan-bad}
can use unicast to announce prefixes for virtual WAN hubs. 
Unicast routing is more deterministic and more likely 
to lead to predictable performance. However, since anycast allows
cloud providers to load-balance across PoPs, split the burden
of flash crowds and denial-of-service attacks, cloud providers 
prefer it over unicast.

\myparab{Interventions from the ISP.}
Today, when performance anomalies like the one in Figure~\ref{fig:vwan-bad}
arise, enterprise customers open \emph{incidents} with cloud providers.
Cloud providers react to these complaints by contacting operators in charge
of the enterprise's upstream ISP. Dialogue between the cloud and ISP
operators can resolve the inflated latency issue if the ISP
changes network configuration to route traffic to the closer point-of-presence.
While this is the common-case resolution, it is time consuming and leads
to lost revenue for both cloud and enterprise clients.
\section{Related Work}
\label{sec:related}
We discuss prior work related to \sysname, and 
set them into context with our contributions.  

\myparab{Overlay networks} have been an active area of research~\cite{ref:ron, ref:akamai, ref:topo_aware_overlay}.
This work has focussed on designing overlay networks to improve 
performance by recovering and re-routing packets in case of outages or congestion. 
Instead, \sysname minimizes connection latency by placing WAN hubs efficiently.
Overlay networks also have different 
architectures for different applications and services such as content delivery networks, 
peer to peer communications \etc~\cite{ref:overlay_foi}. In contrast, virtual
WANs are WAN-as-a-service offering.

\myparab{CDN replica selection.}
Unlike performance-based 
selection of CDN replicas, our focus is on designing virtualized 
WANs that are stable for extended periods of time despite routing dynamics.

\myparab{Performance monitoring in WANs.} \sysname leverages the dynamic latency measurements in its design to predict the 
most suitable hub locations. Research~\cite{ref:can_they_hear_me, ref:odin} has shown that collecting measurements from locations
where clients are located can help in understanding the latency dynamics of wide area networks. While other work~\cite{ref:stable_wan} has shown that
even though the internet performance is heterogeneous, statistical methods can be formulated to quantify the stability of 
network performance. In \sysname we use the latency measurements to formulate an optimal way of deploying hubs for virtualized WAN topologies. 

\myparab{WAN traffic engineering.}
Cloud providers use software-defined
centralized traffic engineering controllers to assign flow 
within their private WANs to maximize their utilization, 
guarantee fairness and prevent congestion~\cite{ref:swan, ref:b4}. 
Bandwidth costs in the context of WANs were considered in Pretium~\cite{ref:pretium} 
and Cascara~\cite{ref:cascara}. \sysname's cost optimization of
virtual WANs solves a different objective with separate
constraints. 
Recent work has proposed a software-defined
edge to manage outbound flows from their networks~\cite{ref:espresso, ref:edgefabric}.
The goal of these efforts is to react to poor
client performance by switching to better performing BGP next hops.

\section{Conclusions}
The current approach to placing virtual WAN hubs, by geographic
proximity, results in unnecessarily high latencies.  We devoped
\sysname, which  designs performance and cost optimal 
virtual WAN topologies. \sysname formulates virtual WAN design
as a mixed integer linear program. The generaed virtual WAN  topologies
can save upto 28\% of the cost of virtual WAN
deployments by placing fewer WAN hubs while achieving the same
service-level objective. \sysname-designed topologies
reduce weighted latency by 26\% for large enterprise clients.

\label{endOfBody}

\label{pg:endofbody}

\bibliographystyle{plain}
\bibliography{main}

\label{pg:endofpaper}
\appendix
\section{Appendices \label{sec:appendix}}
\subsection{Sustained performance benefit from faster PoPs}
\label{sec:appendix_non_default_count}
\begin{figure}[h]
  \centering
      \begin{subfigure}{0.4\textwidth}
           \centering
           \includegraphics[width=\columnwidth]{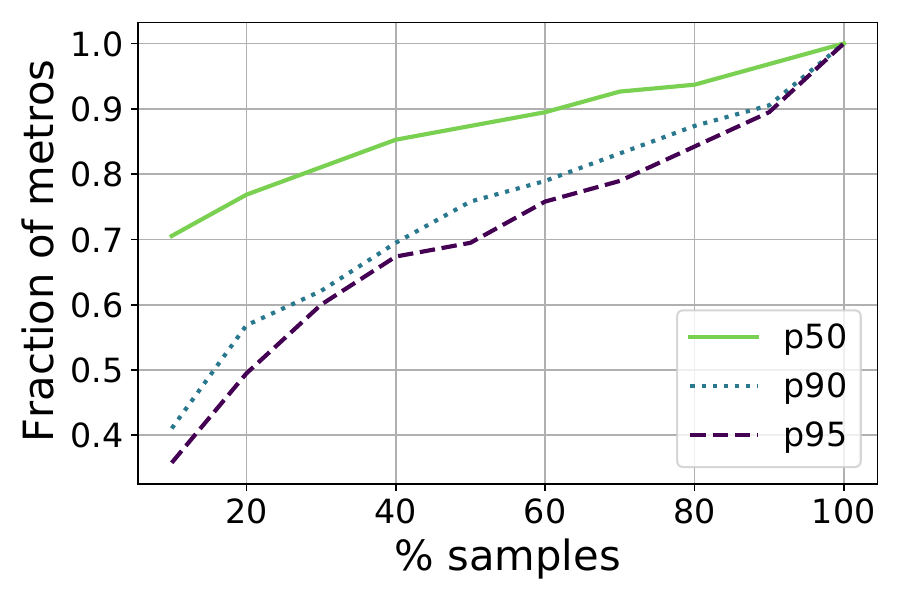}
           \caption{Faster PoP \%  (1 hour)}
           \label{fig:non_default_count_1_hour}
      \end{subfigure}
      \hfill 
      \begin{subfigure}{0.4\textwidth}
           \centering
           \includegraphics[width=\columnwidth]{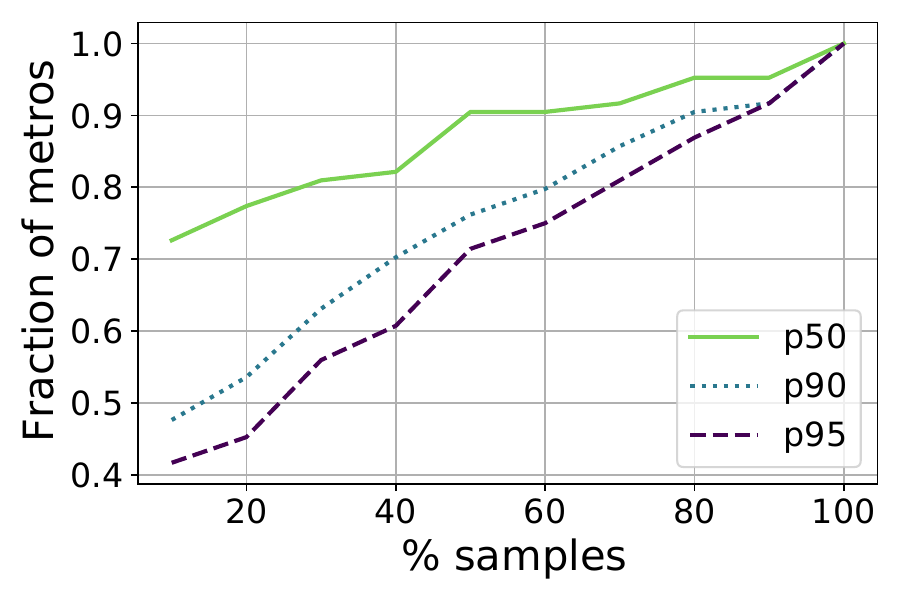}
           \caption{Faster PoP \% (10 hours).}
           \label{fig:non_default_count_10_hour}
       \end{subfigure}
   
        {\vspace{0.15in}}

        \caption{Figures ~\ref{fig:non_default_count_1_hour}-~\ref{fig:non_default_count_10_hour} show performance of faster PoPs remain stable for longer timescales, i.e, 1 hour and 10 hours respectively.}
\end{figure}

\subsection{Quality of Dataset}
\label{sec:appendix_qod}
\begin{figure}[h]
  \centering
  \begin{subfigure}{0.4\textwidth}
    \centering
    \includegraphics[width=\columnwidth]{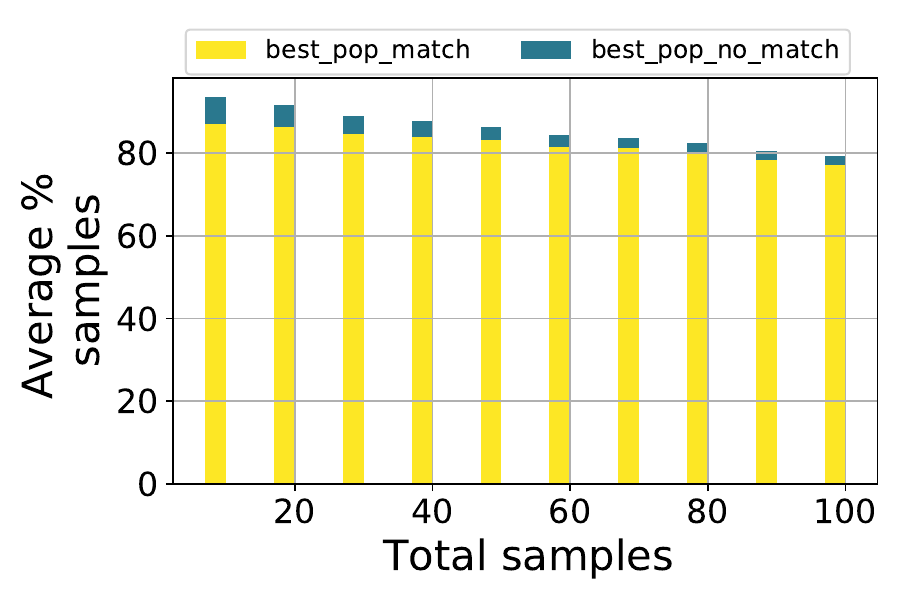}
    \caption{PoP matches over a moving window of 1 day.}
    \label{fig:pop_matches_1_day}
  \end{subfigure}
  \hfill
  \begin{subfigure}{0.4\textwidth}
    \centering
    \includegraphics[width=\columnwidth]{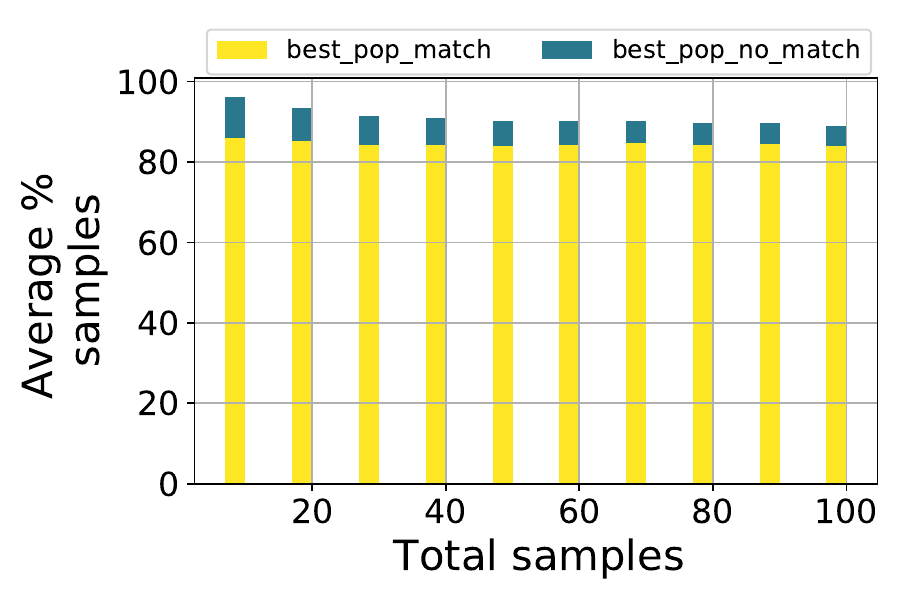}
    \caption{PoP matches over a moving window of 7 days.}
    \label{fig:pop_matches_7_days}
  \end{subfigure}

   {\vspace{0.15in}}
  \caption{Figures ~\ref{fig:pop_matches_1_day}-~\ref{fig:pop_matches_7_days} show the percentage of PoP matches/no-matches
  for a period of 1 day and 7 days with a moving window of 1 day.}
\end{figure}

\subsection{Latency}
\label{sec:appendix_lat}
\begin{figure}[h]
  \centering
  \includegraphics[width=0.44\columnwidth]{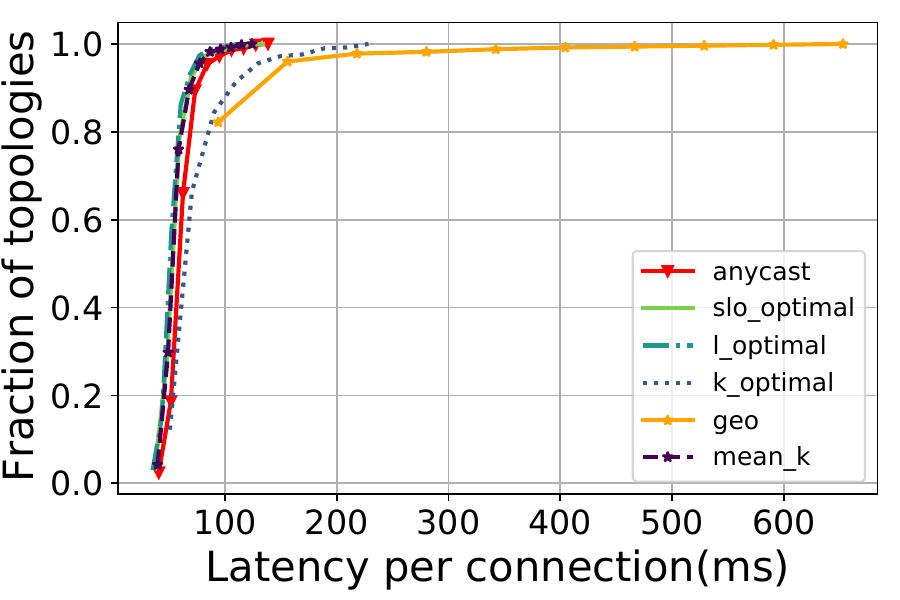}
  \caption{Complete CDF of latency per connection across all the modes.}
  \label{fig:pop_switch_90}
\end{figure}
We show the full graph of the CDF of latency per connection in Figure~\ref{fig:pop_switch_90}. \emph{geo} has a long tail with the maximum latency 
around $\approx$680\emph{ms}. This shows that for a handful of metros where the \emph{geo-default}, and \emph{anycast-default} are different, the 
\emph{geo-default} is much slower.

\subsection{Synthetic Topologies}
\label{sec:synthetic_topologies}
Tables~\ref{tab:topo_eg_1} and ~\ref{tab:topo_eg_2} show two synthetic topologies of size 10. All other topologies have the same structure.

\begin{table}[H]
      \begin{tabular}{|r|r|r|}
      \hline
      \textbf{metro} & \textbf{connections} & \textbf{Default PoP}  \\ \hline
      metro\_15 & 1501 & pop\_1  \\ \hline
      metro\_2 & 36 & pop\_2   \\ \hline
      metro\_34 & 58 & pop\_3  \\ \hline
      metro\_48 & 89 & pop\_4  \\ \hline
      metro\_5 & 4000 & pop\_5   \\ \hline
      metro\_65 & 399 & pop\_2   \\ \hline
      metro\_7 & 408 & pop\_6   \\ \hline
      metro\_89 & 1610 & pop\_7 \\ \hline
      metro\_95 & 519 & pop\_3  \\ \hline
      metro\_10 & 1853 & pop\_8  \\ \hline
      \end{tabular}%
      \caption{Example topology of size 10.}
     \label{tab:topo_eg_1}
\end{table}

\begin{table}[H]
  \begin{tabular}{|r|r|r|}
  \hline
  \textbf{metro} & \textbf{connections} & \textbf{Default PoP}  \\ \hline
  metro\_17 & 584 & pop\_7 \\ \hline
  metro\_35 & 1477 & pop\_4  \\ \hline
  metro\_50 & 213 & pop\_1  \\ \hline
  metro\_110 & 862 & pop\_1  \\ \hline
  metro\_1 & 422 & pop\_5  \\ \hline
  metro\_19 & 1066 & pop\_3  \\ \hline
  metro\_78 & 1093 & pop\_1  \\ \hline
  metro\_99 & 364 & pop\_2 \\ \hline
  metro\_21 & 2613 & pop\_9   \\ \hline
  metro\_6 & 30 & pop\_10  \\ \hline
  \end{tabular}%
  \caption{Example topology of size 10.}
  \label{tab:topo_eg_2}
\end{table}

\subsection{Pareto Frontier}
\label{sec:appendix_pareto}
\begin{figure*}[t]
  \centering
  \begin{subfigure}{0.32\textwidth}
    \centering
    \includegraphics[width=\columnwidth]{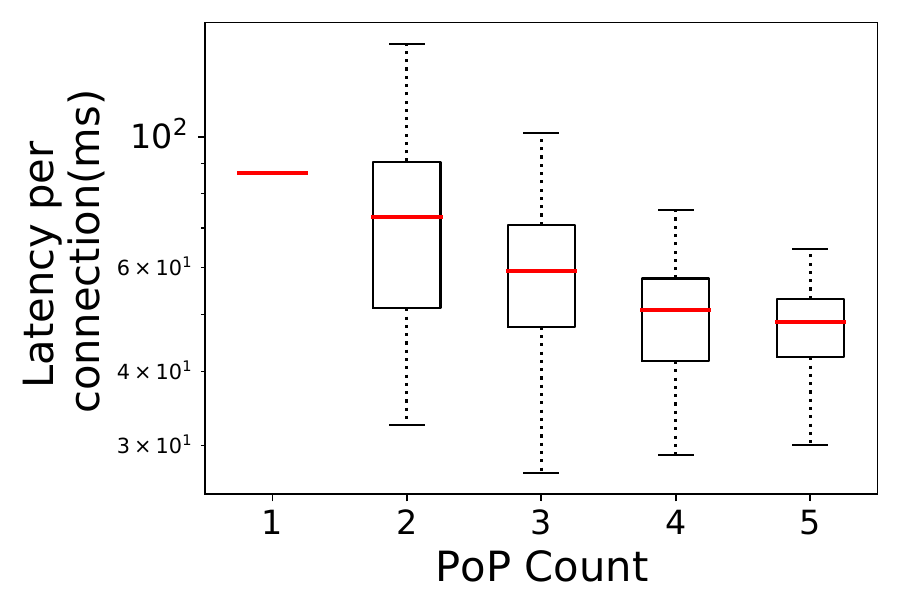}
    \caption{Distribution of latency per connection on the Pareto Frontier for Topology of size 5.}
    \label{fig:pareto_5}
  \end{subfigure}
  \hfill
  \begin{subfigure}{0.32\textwidth}
    \centering
    \includegraphics[width=\columnwidth]{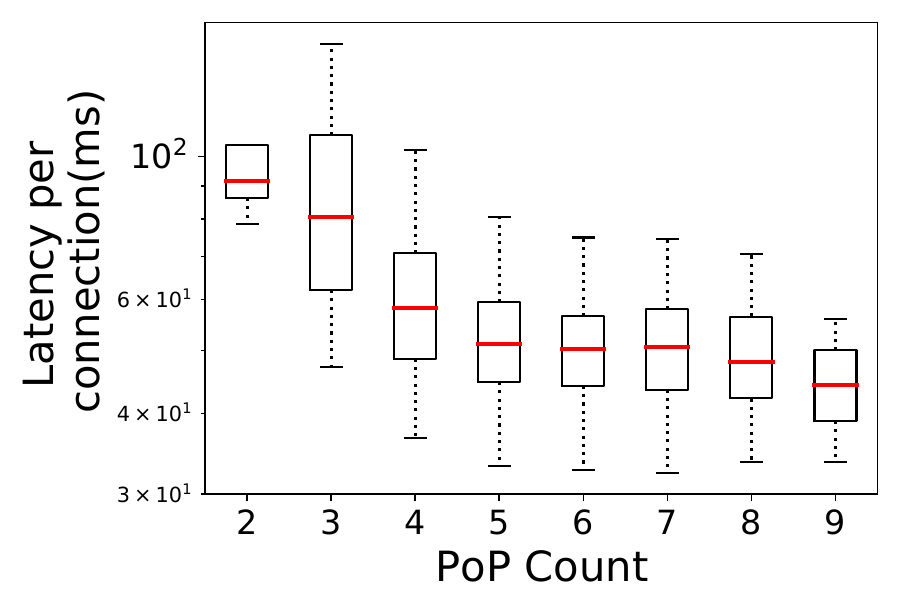}
    \caption{Distribution of latency per connection on the Pareto Frontier for Topology of size 10.}
    \label{fig:pareto_10}
  \end{subfigure} 
  \hfill
  \begin{subfigure}{0.32\textwidth}
    \centering
    \includegraphics[width=\columnwidth]{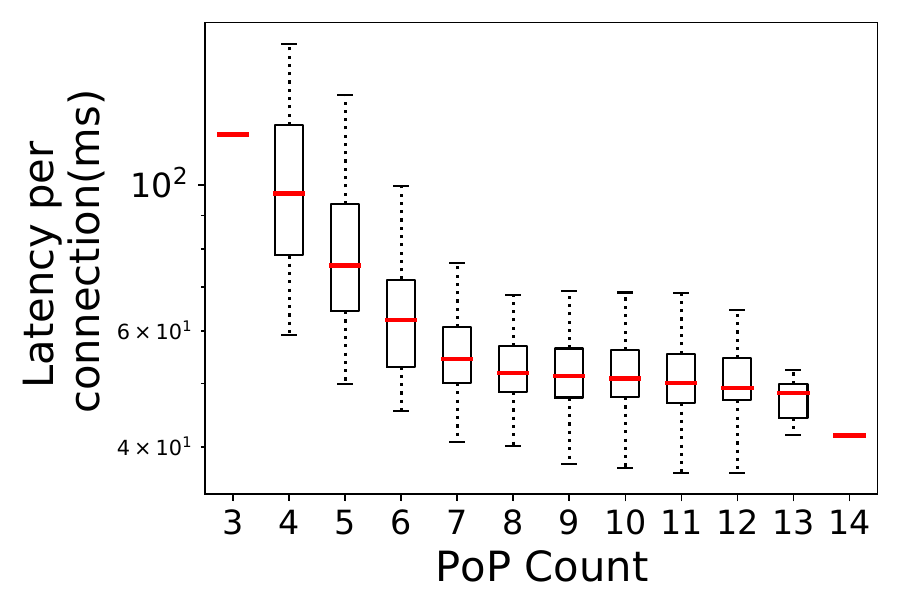}
    \caption{Distribution of latency per connection on the Pareto Frontier for Topology of size 25.}
    \label{fig:pareto_25}
  \end{subfigure} 
   
   {\vspace{0.25in}}
  \caption{Figures~\ref{fig:pareto_5}-~\ref{fig:pareto_25} show the distribution of latency on the Pareto Frontier of performance vs. cost.
  Latency benefits diminish after a particular PoP count across all topology sizes.}
\end{figure*}

We show the latency distribution at all possible PoP counts for topologies of sizes, 
5, 10 and 25 in Figures~\ref{fig:pareto_5}-~\ref{fig:pareto_25}.

\begin{figure*}[h]
  \centering
  \begin{subfigure}{0.32\textwidth}
    \centering
    \includegraphics[width=\columnwidth]{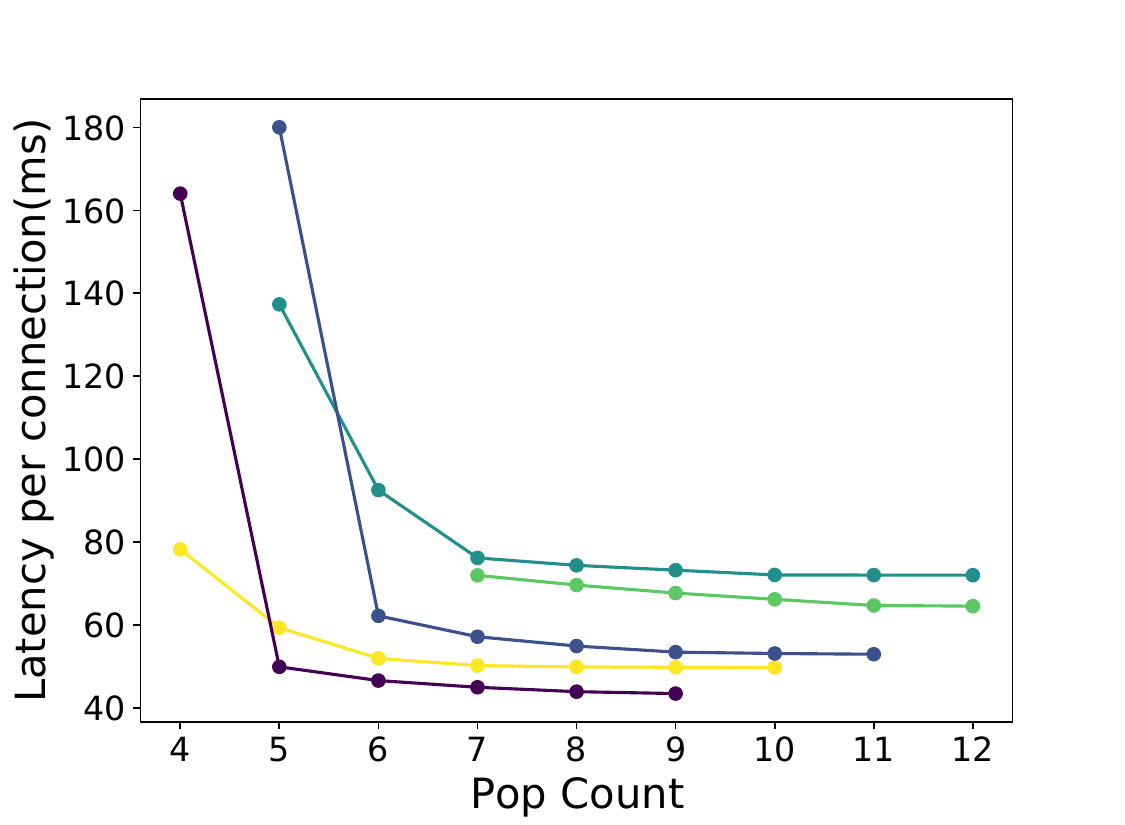}
    \caption{Topology 25.}
    \label{fig:pareto_curve_25}
  \end{subfigure}
  \hfill
  \begin{subfigure}{0.32\textwidth}
    \centering
    \includegraphics[width=\columnwidth]{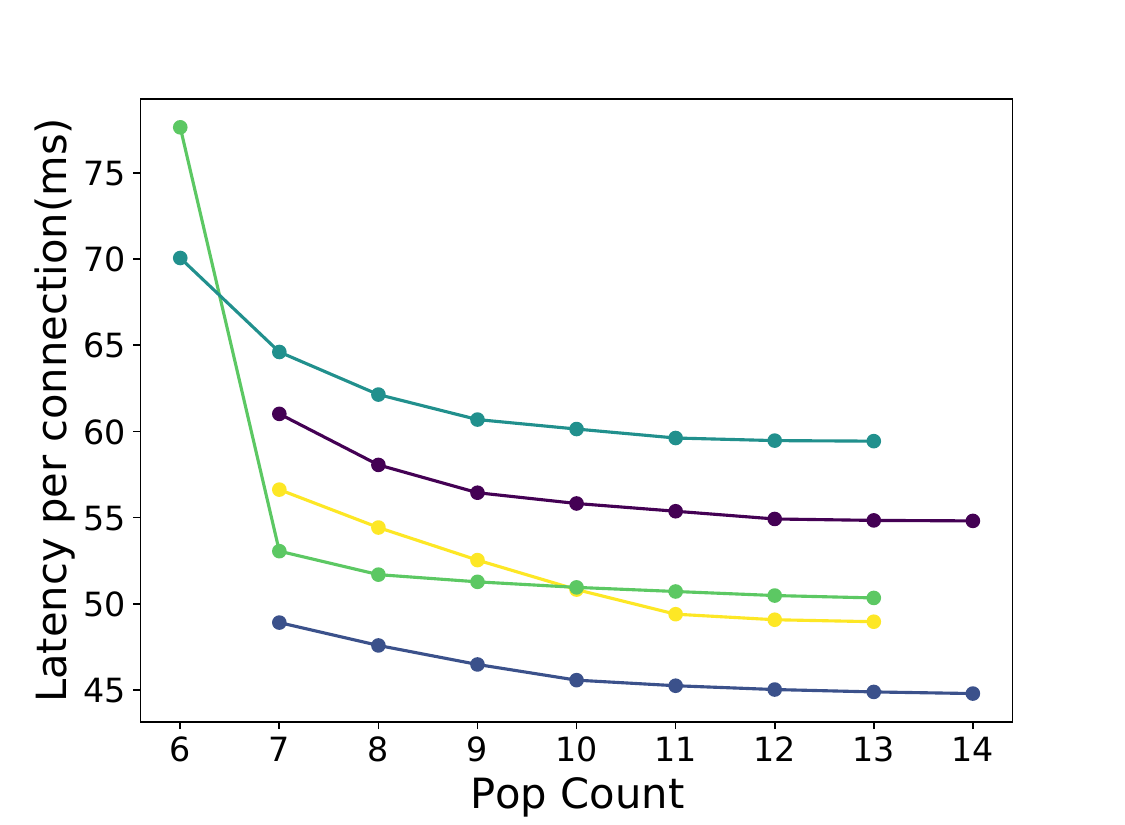}
    \caption{Topology 50.}
    \label{fig:pareto__curve_50}
  \end{subfigure} 
  \hfill
  \begin{subfigure}{0.32\textwidth}
    \centering
    \includegraphics[width=\columnwidth]{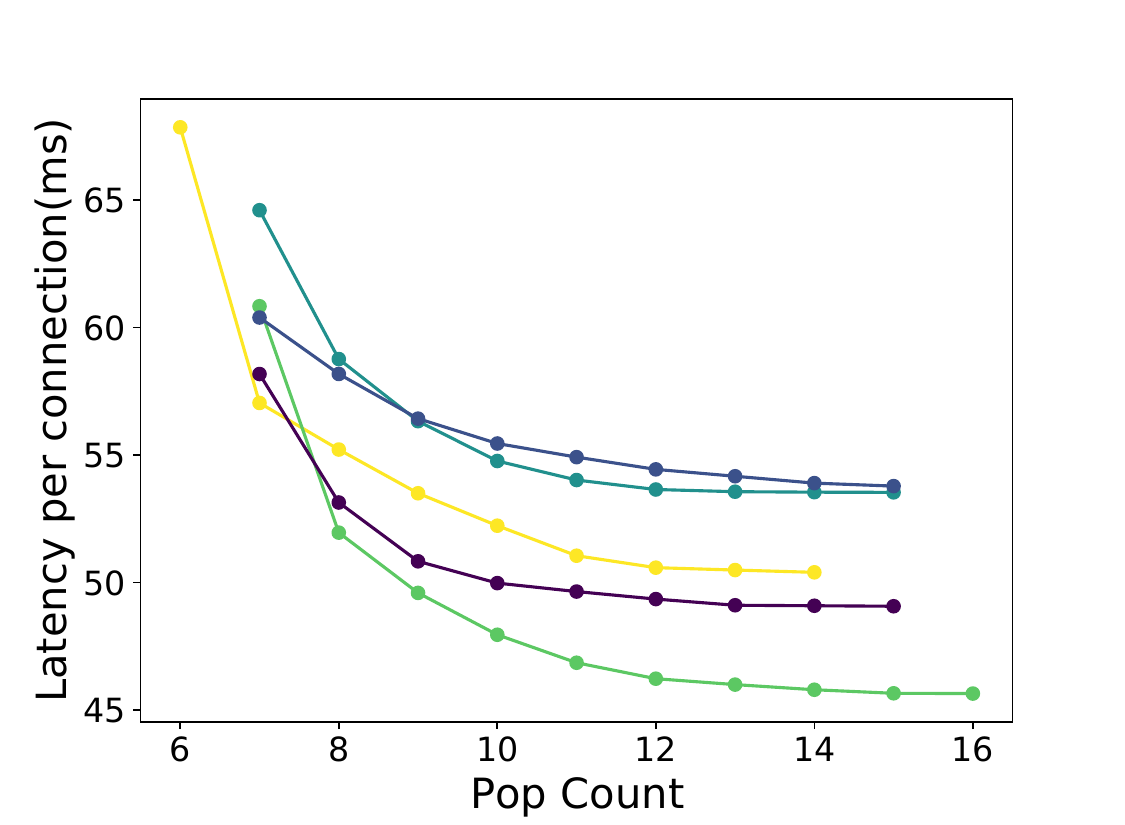}
    \caption{Topology 75.}
    \label{fig:pareto_curve_75}
  \end{subfigure} 

   {\vspace{0.25in}}
  \caption{Figures~\ref{fig:pareto_curve_25}-~\ref{fig:pareto_curve_75} show the trends in latency per connection with different 
  PoP counts for 5 topologies of each size. All sizes show when the curve reaches the knee, \ie \stt{mean\_k} the latency benefits start to diminish.}
\end{figure*} 


\begin{figure*}[h]
  \centering
  \begin{subfigure}{0.30\textwidth}
    \centering
    \includegraphics[width=\columnwidth]{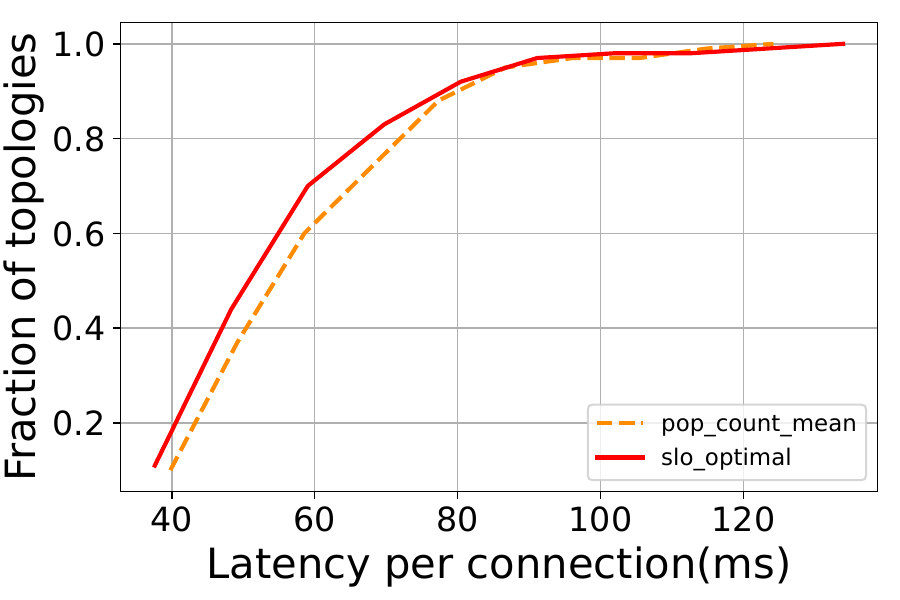}
    
    \caption{Topology 5.}
    \label{fig:all_pops_lat_5}
  \end{subfigure}
  \hfill
  \begin{subfigure}{0.30\textwidth}
    \centering
    \includegraphics[width=\columnwidth]{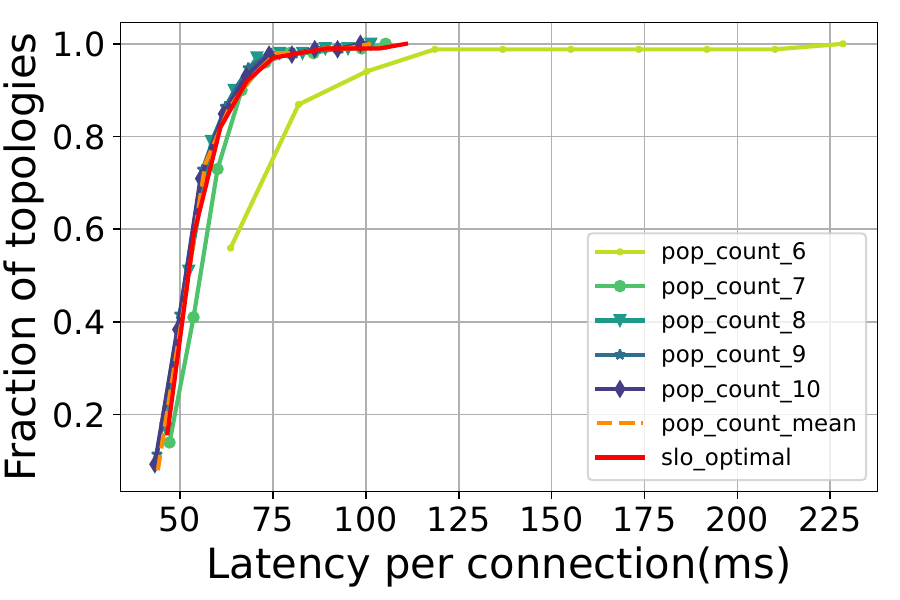}
    \caption{Topology 25.}
    \label{fig:all_pops_lat_25}
  \end{subfigure} 
  \hfill
  \begin{subfigure}{0.30\textwidth}
    \centering
    \includegraphics[width=\columnwidth]{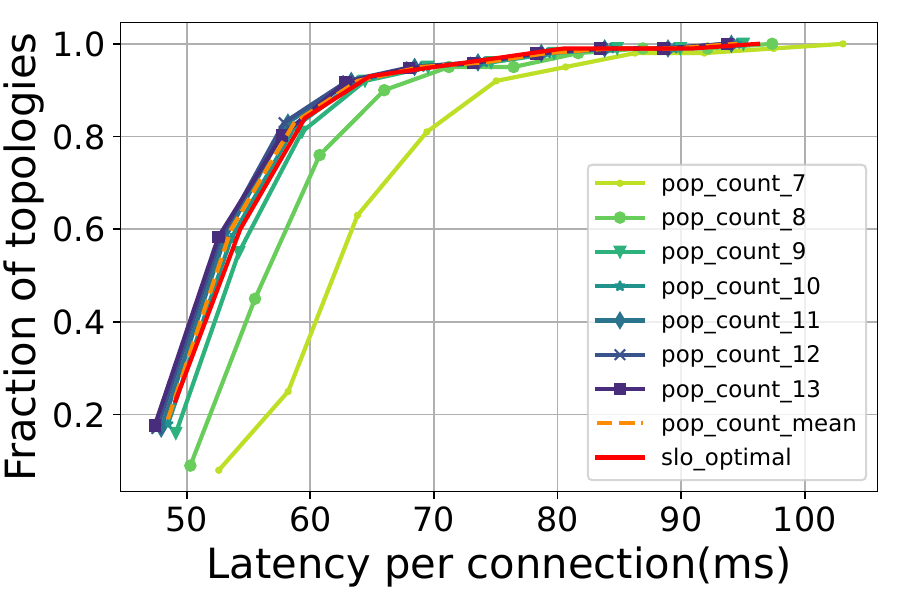}
    \caption{Topology 50.}
    \label{fig:all_pops_lat_50}
  \end{subfigure} 
   {\vspace{0.25in}}
  \caption{Figures~\ref{fig:all_pops_lat_5}-~\ref{fig:all_pops_lat_50} the latency distribution at each PoP count. We only include
  PoP count for topologies that gave a feasible solution was 80 or more.}
\end{figure*}

\subsection{Pareto Efficiency Curve examples}
\label{sec:appendix_pareto_curve}
We show the trends of latency per connection at each PoP count for five individual topologies of sizes, 25, 50, and 75 in  
Figures~\ref{fig:pareto_curve_25}-~\ref{fig:pareto_curve_75}.

\subsection{All PoP Counts Latencies}
\label{sec:appendix_lat_dist}
We show the latency distribution across different topology sizes with all possible PoP counts in  
Figures~\ref{fig:all_pops_lat_5}-~\ref{fig:all_pops_lat_50}.


\begin{figure}[t]
  \centering
  \begin{subfigure}[t]{0.44\columnwidth}
    \includegraphics[width=\columnwidth]{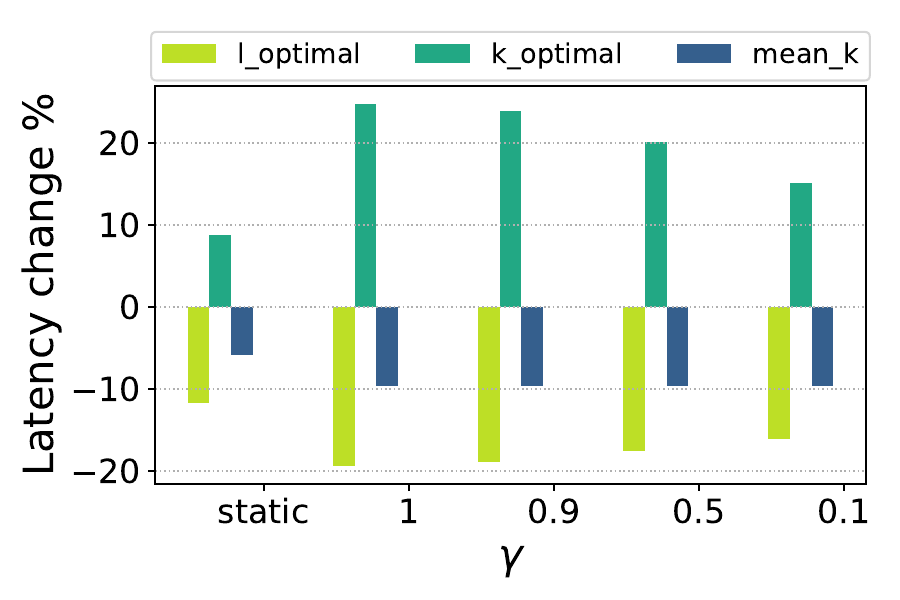}
    \caption{Latency changes.}
    {\vspace{0.15in}}
    \label{fig:lat_maintain_with_geo_1_day}
  \end{subfigure}
  \hfill
  \begin{subfigure}[t]{0.44\columnwidth}
    \includegraphics[width=\columnwidth]{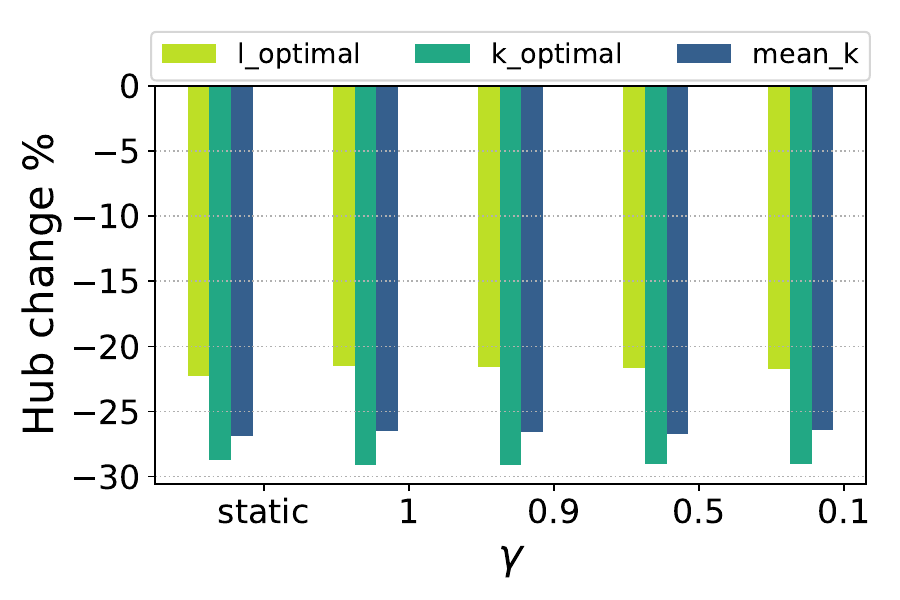}
    \caption{Hub count changes.}
    \label{fig:hubs_maintain_with_geo_1_day}
   \end{subfigure}

   {\vspace{0.2in}}
  \caption{~\ref{fig:lat_maintain_with_geo_1_day}-~\ref{fig:hubs_maintain_with_geo_1_day} shows the latency, and hub count with different 
            smoothing variations with geo baseline on a moving window of 1-day.
  }
\end{figure}

\begin{figure}[t]
  \centering
  \begin{subfigure}[t]{0.44\columnwidth}
    \includegraphics[width=\columnwidth]{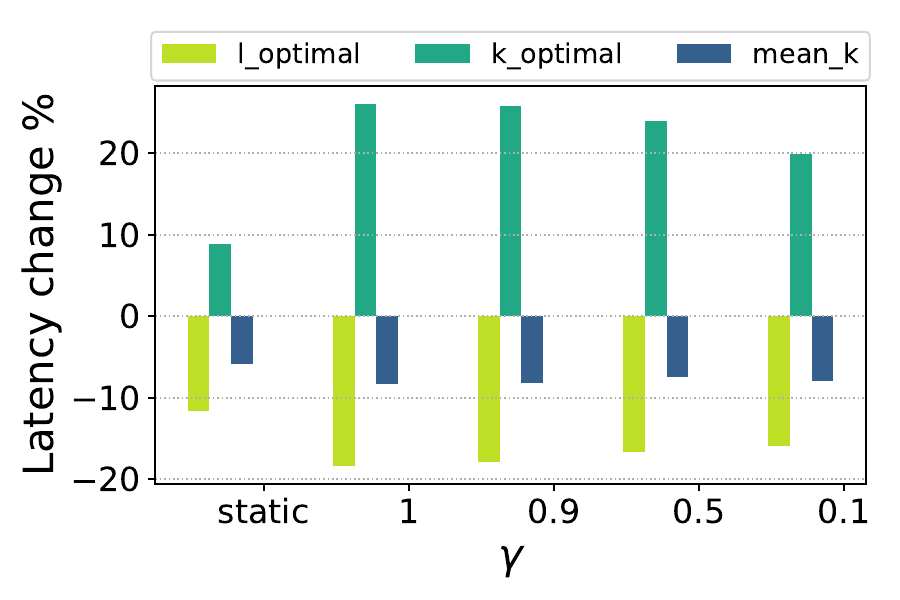}
    \caption{Latency changes.}
    {\vspace{0.15in}}
    \label{fig:lat_maintain_with_geo_2_days}
  \end{subfigure}
  \hfill
  \begin{subfigure}[t]{0.44\columnwidth}
    \includegraphics[width=\columnwidth]{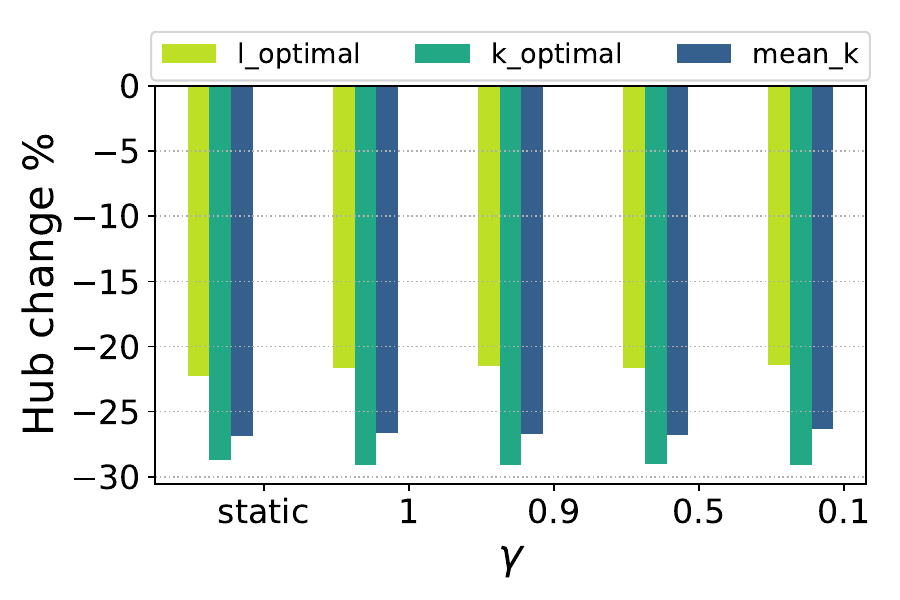}
    \caption{Hub count changes.}
    \label{fig:hubs_maintain_with_geo_2_days}
   \end{subfigure} 

   {\vspace{0.2in}}
  \caption{~\ref{fig:lat_maintain_with_geo_2_days}-~\ref{fig:hubs_maintain_with_geo_2_days} shows the latency, and hub count with
             different smoothing variations with geo baseline on 2-days window.}
\end{figure}

\subsection{Maintenance}
\label{sec:appendix_maintain}

We show the maintenance results for 1-day in  
Figures~\ref{fig:lat_maintain_with_geo_1_day}-~\ref{fig:hubs_maintain_with_geo_1_day}. The results for 2-days windows are
shown in Figures~\ref{fig:lat_maintain_with_geo_2_days}-~\ref{fig:hubs_maintain_with_geo_2_days}

\end{document}